\documentclass[12pt]{article}
\usepackage{graphicx}
\usepackage{multicol}
\font\tenrsfs=rsfs10 at 12pt
\font\sevenrsfs=rsfs7
\font\fiversfs=rsfs5
\newfam\rsfsfam
\textfont\rsfsfam=\tenrsfs
\scriptfont\rsfsfam=\sevenrsfs
\scriptscriptfont\rsfsfam=\fiversfs
\def\mathscr#1{{\fam\rsfsfam\relax#1}}
\def\Lag{\mathscr{L}}

\newcommand{\fig}[1]{~{\rm \ref{fig:#1}}}

\newcommand{\GeV}{\,{\rm GeV}}

\newcommand{\TeV}{\,{\rm TeV}}

\newcommand{\yuk}{{Y}}
\newcommand{\diagyuk}{{\lambda}}
\newcommand{\hatdyuk}{{\hat \lambda}}
\newcommand{\realV}{V}
\newcommand{\hatV}{{\hat V}}
\newcommand{\hatFC}{{\hat \lambda_{\rm FC}}}

\newcommand{\be}{\begin{equation}}
\newcommand{\ee}{\end{equation}}
\newcommand{\bea}{\begin{eqnarray}}
\newcommand{\eea}{\end{eqnarray}}
\newcommand{\beq}{\begin{equation}}
\newcommand{\eeq}{\end{equation}}
\newcommand{\beqa}{\begin{eqnarray}}
\newcommand{\eeqa}{\end{eqnarray}}
\newcommand{\ba}{\begin{array}}
\newcommand{\ea}{\end{array}}

\newcommand{\cL}{\Lag}
\newcommand{\cH}{\mathscr{H}}
\newcommand{\cA}{{\cal A}}
\newcommand{\cO}{{\cal O}}

\newcommand{\no}{\nonumber}

\newcommand{\sw}{\sin^2\theta_{\rm W}}
\newcommand{\BR}{{\cal B}}

\newcommand{\muw}{\mu_{\scriptscriptstyle W}}

\newcommand{\mg}{m_{\tilde g}}
\newcommand{\msqL}{m_{{\tilde q}_L}}

\newcommand{\msdR}{m_{{\tilde d}_R}}

\newcommand{\msuR}{m_{{\tilde u}_R}}

\def\identity{1 \hspace{-.085cm}{\rm l}}
\def\circa#1{\,\raise.3ex\hbox{$#1$\kern-.75em\lower1ex\hbox{$\sim$}}\,}

\def\Black{}

\def\npb#1#2#3{    {\it Nucl. Phys. }{\bf B #1} (#2) #3}
\def\plb#1#2#3{    {\it Phys. Lett. }{\bf B #1} (#2) #3}

\def\prd#1#2#3{    {\it Phys. Rev. }{\bf D #1} (#2) #3}

\def\prep#1#2#3{   {\it Phys. Rep. }{\bf #1} (#2) #3}
\def\prl#1#2#3{    {\it Phys. Rev. Lett. }{\bf #1} (#2) #3}

\def\rmp#1#2#3{    {\it Rev. Mod. Phys. }{\bf #1} (#2) #3}
\def\zpc#1#2#3{    {\it Z. Phys. }{\bf C #1} (#2) #3}
\def\ijmpa#1#2#3{  {\it Int. J. Mod. Phys. }{\bf A #1} (#2) #3}

\def\epjc#1#2#3{   {\it Eur. Phys. J. }{\bf C #1} (#2) #3}
\def\ibid#1#2#3{   {\it ibid. }{\bf #1} (#2) #3}
\def\jhep#1#2#3{   {\it JHEP  }{\bf #1} (#2) #3}

\begin{document}
 \centerline{hep-ph/0207036 \hfill CERN--TH/2002--147\hfill IFUP--TH/2002--17}
\vspace{8mm}
\Black
\vspace{0.5cm}
\centerline{\Large\bf Minimal Flavour Violation:}
\vspace{2mm}
\centerline{\Large\bf an effective field theory approach}
\Black
 \vspace{0.8cm}
    \centerline{\bf G. D'Ambrosio,$^{a,b}$ G.F. Giudice,$^{a}$ G. Isidori,$^{a,c}$
                 A. Strumia$^{a,d}$}
 \vspace{0.8cm}
\centerline{\em ${}^a$Theoretical Physics Division, CERN, CH-1211 Gen\`eve 23, Switzerland}
 \vspace{0.2cm}
\centerline{\em ${}^b$INFN, Sezione di Napoli, I-80126 Napoli, Italy}
 \vspace{0.2cm}
\centerline{\em ${}^c$INFN, Laboratori Nazionali di Frascati, I-00044 Frascati, Italy}
 \vspace{0.2cm}
\centerline{\em ${}^d$Dipartimento di Fisica
dell'Universit\`a di Pisa and INFN}
 \vspace{1cm}
\centerline{\large\bf Abstract}
\begin{quote}
\indent
We present a general analysis of extensions of the Standard Model
which satisfy the criterion of Minimal Flavour Violation (MFV). We define this
general framework by constructing a low-energy effective theory containing
the Standard Model fields, with one or two Higgs doublets and, as the only
source of ${\rm SU}(3)^5$ flavour symmetry breaking, the background values
of fields transforming under the flavour
group as the ordinary Yukawa couplings. We analyse present bounds on the
effective scale of dimension-six operators, which range between 1 and
10 TeV, with the most stringent constraints imposed by $B\to X_s \gamma$.
In this class of theories, it is possible to relate predictions for FCNC
processes in $B$ physics to those in $K$ physics. We compare the sensitivity of
various experimental searches in probing the hypothesis of MFV.
Within the two-Higgs-doublet scenario, we develop a general procedure
to obtain all $\tan\beta$-enhanced Higgs-mediated FCNC amplitudes, discussing
in particular their impact in $B\to\ell^+\ell^-$,  $\Delta M_B$ and
$B\to X_s \gamma$. As a byproduct, we derive some two-loop $\tan\beta$-enhanced
supersymmetric contributions to $B\to X_s \gamma$ previously unknown.
\end{quote}
\vspace{5mm}

\renewcommand{\thefootnote}{\arabic{footnote}}

\section{Introduction}
The Standard Model (SM) is a successful effective theory of particle
interactions valid up to some still undetermined cut-off
energy scale $\Lambda$. The goal of the search for physics beyond the
SM is to find evidence for effects that are present in the theory with a finite
value of $\Lambda$, but disappear in the limit $\Lambda \to \infty$.
The theoretical argument based on a natural solution of the hierarchy
problem requires that $\Lambda$ should not exceed few TeV, in order
to stabilize the Higgs mass parameter. Experimental searches mostly provide
lower bounds on $\Lambda$. Evidence for neutrino masses and limits on
the proton lifetime push the effective scale of baryon and lepton number violating
interactions close to the GUT scale. Limits on electron and neutron electric
dipole moments require that the scale of (flavour-conserving) CP violation
is larger than $10^7$~GeV, while the $K^0$--${\bar K}^0$ mass difference
sets a lower bound of about $10^6$~GeV on the scale of $\Delta S=2$ flavour
transitions. Generic contact interactions that preserve all selection
rules are limited by LEP data, and their effective scales have bounds
in the 1--10~TeV range.

If we insist with the theoretical prejudice that new physics has to emerge
in the TeV region, then we have to conclude that the new theory is highly
non-generic and we should use the lower limits on $\Lambda$ to
constrain its form. In particular, $B$, $L$, and CP have to be approximate
symmetries of the new theory at the TeV scale. The case of the flavour
symmetry is less straightforward. Indeed the ${\rm U}(3)^5$ flavour symmetry
of the SM gauge interactions is already broken by Yukawa couplings.
Therefore, it does not seem very plausible to impose
that the new interactions respect flavour, a symmetry
which is not realized in the SM,
{\it i.e.} in the low-energy limit of the new theory. On the other hand,
generic flavour-violating interactions at $\Lambda \simeq$~TeV are
experimentally excluded.

The most reasonable solution to this impasse is to impose that
the effective theory respect what we call {\it Minimal Flavour Violation}
(MFV). Although more rigorously defined in section~\ref{secmfv}, MFV essentially
requires that all  flavour and CP-violating interactions are linked to the
known structure of Yukawa couplings.
This concept is a familiar one in the case of supersymmetry,
where this hypothesis is often adopted.
In this paper we take a more general point of view:
we show how the MFV hypothesis can be consistently
defined independently of the structure of the new-physics model,
we describe the generic form of an effective
theory obeying this hypothesis, and derive the present constraints
on the characteristic scale $\Lambda$. As shown in section~\ref{sect:bounds},
we find that the constraints vary
from 1~TeV to about 10 TeV, with the strongest bound coming from $B \to
X_s \gamma$.

In the context of MFV it is possible to relate  various  flavour-changing
neutral current (FCNC) processes. Not only is this possible within the
$B$ or $K$ systems, but one can also relate predictions for
$B$ processes to $K$ rare decays. This can be achieved because, as shown in
section~\ref{secmfv}, MFV implies that all flavour-changing effective
operators are proportional to the same non-diagonal structure. In turn,
this is a consequence of the top Yukawa coupling being much larger than
all other Yukawa couplings.

An interesting novelty emerges in models in which the low-energy limit
is not simply described by the SM, but it is enlarged to contain two
Higgs doublets. If $\tan\beta$ (the ratio between the two vacuum
expectation values) is large, then the bottom Yukawa coupling can become
comparable to the top-quark coupling and the operators of the MFV effective
theory contain a second non-negligible flavour-violating structure.
Particularly interesting is the case in which the masses of the new
Higgs bosons are smaller than $\Lambda$, since
the new effects are perturbatively computable. In section~\ref{sect:2HD} we derive
the form of the currents coupled to the neutral and charged Higgs bosons,
including all terms enhanced by $\tan\beta$ and in section~\ref{sect:2HD_FCNC} we
discuss their phenomenological implications.

We believe that the description of flavour violations in terms of an
effective theory is a useful tool to analyse future data on $B$ and $K$
mesons. When deviations from the SM predictions are observed, the comparison
among various processes will indicate if they can be described in the general
framework of MFV, or if they require new flavour structures. This will be
a crucial hint to identify the correct theory valid above the cut-off
scale $\Lambda$.

In the case of supersymmetry, the MFV hypothesis is valid when the
scalar-mass soft terms are universal and the trilinear soft terms are
proportional to Yukawa couplings, at an arbitrary high-energy scale. Then
the physical squark masses are not equal, but the induced flavour violation
is described in terms of the usual CKM parameters. In this case, the
coefficients of the MFV operators described in section~\ref{secop} can be
perturbatively computed in terms of supersymmetric masses. However, given the
large number of free parameters in a supersymmetric model, the
effective-theory description can still be useful as a bookkeeping device.
The concrete meaning of MFV in supersymmetric models is discussed 
in section~\ref{sect:examples}. Moreover,
in  section~\ref{sect:2HD_FCNC} we show how the effective theory 
approach provides a simple systematic tool to identify
$\tan\beta$-enhanced Higgs-mediated FCNC amplitudes and,
using it, we identify a contribution to
$B\to X_s\gamma$ which so far has not been
discussed in the literature.

On the other hand, the effective-theory language becomes compulsory
to address the flavour problem in models with
non-perturbative interactions at energies $\Lambda$,
like in scenarios with low-energy quantum-gravity scale.
In section~\ref{sect:examples} we shall briefly discuss
the natural size of the effective operators
expected in some of these models.

\section{Minimal Flavour Violation}
\label{secmfv}

The SM fermions consist of three families with two
${\rm SU}(2)_L$ doublets ($Q_L$ and $L_L$) and three ${\rm SU}(2)_L$
singlets ($U_R$, $D_R$ and $E_R$).
The largest group of unitary field transformations that commutes with the gauge
group is ${\rm U}(3)^5$~\cite{Georgi}.  This can be decomposed as
\beq
G_F \equiv {\rm SU}(3)^3_q \otimes  {\rm SU}(3)^2_\ell
\otimes  {\rm U}(1)_B \otimes {\rm U}(1)_L \otimes {\rm U}(1)_Y \otimes {\rm U}(1)_{\rm PQ} \otimes {\rm U}(1)_{E_R}~,
\eeq
where
\beqa
{\rm SU}(3)^3_q     &=& {\rm SU}(3)_{Q_L}\otimes {\rm SU}(3)_{U_R} \otimes {\rm SU}(3)_{D_R}~,  \\ \no
{\rm SU}(3)^2_\ell  &=&  {\rm SU}(3)_{L_L} \otimes {\rm SU}(3)_{E_R}~.
\eeqa

Out of the five ${\rm U}(1)$ charges, three can be identified with
baryon ($B$) and lepton ($L$) numbers and hypercharge ($Y$),
which are respected by Yukawa interactions.
The two remaining ${\rm U}(1)$ groups can be identified with the
Peccei-Quinn symmetry of two-Higgs-doublet models~\cite{PQ}
and with a global rotation of a single ${\rm SU}(2)_L$ singlet.
Rearranging these two groups, we denote by ${\rm U}(1)_{\rm PQ}$
a rotation which affects only and in the same way
$D_R$ and $E_R$, and by ${\rm U}(1)_{E_R}$ a rotation of
$E_R$ only. The breaking of these two ${\rm U}(1)$ play
an important role in flavour dynamics
in models with more than one Higgs doublet, where the source of the breaking
can be controlled by fields different than those that generate the Yukawa
couplings. This case will be analysed in section~\ref{sect:2HD}.

In the SM the Yukawa interactions break the symmetry group
${\rm SU}(3)^3_q \otimes  {\rm SU}(3)^2_\ell \otimes {\rm U}(1)_{\rm PQ}
\otimes {\rm U}(1)_{E_R}$. We can formally recover flavour invariance
by introducing dimensionless auxiliary fields $\yuk_U$, $\yuk_D$, and
$\yuk_E$ transforming under ${\rm SU}(3)^3_q \otimes  {\rm SU}(3)^2_\ell$
as
\beq
\yuk_U \sim (3, \bar 3,1)_{{\rm SU}(3)^3_q}~,\qquad
\yuk_D \sim (3, 1, \bar 3)_{{\rm SU}(3)^3_q}~,\qquad
\yuk_E \sim (3, \bar 3)_{{\rm SU}(3)^2_\ell}~.
\eeq
This allows the appearance of Yukawa interactions, consistently with
the flavour symmetry
\beq
\cL  =   {\bar Q}_L \yuk_D D_R  H
+ {\bar Q}_L {\yuk_U} U_R  H_c
+ {\bar L}_L {\yuk_E} E_R  H {\rm ~+~h.c.}~,
\label{eq:LY}
\eeq
where  $H_c = i\tau_2 H^*$ and
$\langle H^\dagger H \rangle = v^2/2$, with $v=246\GeV$.
Notice that eq.~(\ref{eq:LY}) describes the most general coupling
of the fields $\yuk$ to renormalizable SM operators. Indeed, couplings
of $\yuk$ with the kinetic
terms of quarks and leptons can be eliminated with a redefinition of
the fermionic fields, and Yukawa interactions with more $\yuk$ insertions
can be absorbed in a redefinition of the fields $\yuk$.

Using the ${\rm SU}(3)^3_q \otimes  {\rm SU}(3)^2_\ell$
symmetry, we can rotate the background values of the auxiliary fields $\yuk$
such that
\beq
\yuk_D = \diagyuk_d~, \qquad
\yuk_L = \diagyuk_\ell~, \qquad
\yuk_U =  V^\dagger \diagyuk_u~,
\label{eq:d-basis}
\eeq
where $\diagyuk$ are diagonal matrices
and $V$ is the CKM matrix.

We define that an effective theory satisfies the criterion of
{\it Minimal Flavour Violation} if all higher-dimensional operators,
constructed from SM and $\yuk$ fields, are invariant under CP and (formally)
under the flavour group $G_F$. In other words, MFV requires that the
dynamics of flavour violation is completely determined by the structure
of the ordinary Yukawa couplings. 
In particular, all CP violation originates from the CKM phase.

It is easy to realize that, since the SM Yukawa couplings for all fermions
except the top are small, the only relevant non-diagonal structure is
obtained by contracting two $\yuk_U$, transforming as $(8,1,1)$.
For later convenience we define
\beq\label{eq:FC}
(\lambda_{\rm FC})_{ij} = \left\{ \ba{ll} \left( \yuk_U \yuk_U^\dagger \right)_{ij}
\approx \lambda_t^2  V^*_{3i} V_{3j}~ &\qquad i \not= j~, \\
0 &\qquad i = j~. \ea \right.
\eeq
The off-diagonal component of a generic polynomial $P(\yuk_U \yuk_U^\dagger )$ 
is (approximately) proportional to $\lambda_{\rm FC}$. 
Therefore, $\lambda_{\rm FC}$
is the effective coupling governing all FCNC processes with
external down-type quarks.

\section{The MFV dimension-six operators}
\label{secop}

In this section, we want to construct all possible dimension-six FCNC
operators, in the context of an effective theory with MFV.
For processes with external down-type quarks,
three basic bilinear FCNC structures can
be identified:
\beq
{\bar Q}_L \yuk_U \yuk_U^\dagger Q_L~,  \qquad
{\bar D}_R \yuk_D^\dagger \yuk_U \yuk_U^\dagger Q_L~, \qquad
{\bar D}_R \yuk_D^\dagger \yuk_U \yuk_U^\dagger \yuk_D D_R~.
\label{eq:3FCNC}
\eeq
Expanding
in powers of off-diagonal CKM matrix elements
and in powers of small Yukawa couplings,
it is easy to realize that the only relevant
bilinear FCNC structures are
\beq
{\bar Q}_L \lambda_{\rm FC}  Q_L\qquad {\rm and} \qquad
{\bar D}_R \diagyuk_d \lambda_{\rm FC} Q_L~.
\label{eq:str_1}
\eeq
Although
suppressed by $\diagyuk_d$, the $LR$ term has
to be kept because of its unique
${\rm SU}(2)_L$ transformation property.
Similarly to the SM case, FCNC involving external
up-type quarks are absolutely negligible in this
context, because of the smallness of down-type
Yukawa couplings.

We are now ready to build the complete basis of gauge-invariant
dimension-six FCNC operators
(relevant for processes with external down-type quarks),
obtained by breaking $G_F$ only by means of the background values
of $\yuk$. Since we are interestesd also in amplitudes mediated by  
off-shell gauge bosons, we start writing a basis where we use 
the equations of motion (to reduce the number 
of independent terms) only on the fermion fields. 
The operators can be classified as follows:
\begin{itemize}
\item{$\underline{\Delta F=2.}$} Non-negligible $\Delta F=2$
structures are of the type $({\bar Q}_L \lambda_{\rm FC}  \gamma^\mu X^a  Q_L)^2$
where $X^a$ denotes either the identity or a generator of ${\rm SU}(3)_c
\otimes {\rm SU}(2)_L$.
In principle, these lead to four independent operators. However,
once we restrict the attention to down-type quarks, by use of Fiertz
identities all these structures turn out to be equivalent and we are
left with a single independent term, which we choose to be
\beq
{\cal O}_0 = \frac{1}{2}\left({\bar Q}_L \lambda_{\rm FC} \gamma_\mu  Q_L \right)^2~.
\label{eq:O0}
\eeq

\item{$\underline{\Delta F=1\ \mbox{Higgs field}.}$} Neglecting terms
which are suppressed by quark masses via the equations of motion,
we are left with the following two operators
\beq
{\cal O}_{H1} = i \left( {\bar Q}_L \lambda_{\rm FC} \gamma_\mu Q_L \right)
 H^\dagger D_\mu H~, \qquad
{\cal O}_{H2} = i \left( {\bar Q}_L \lambda_{\rm FC} \tau^a \gamma_\mu Q_L\right)
 H^\dagger \tau^a D_\mu H~.
\eeq

\item{$\underline{\Delta F=1\ \mbox{gauge fields}.}$}
The couplings of the currents (\ref{eq:str_1}) to the gluon field are
\beq
{\cal O}_{G1} =   H^\dagger \left( {\bar D}_R  \diagyuk_d \lambda_{\rm FC} \sigma_{\mu\nu}
  T^a  Q_L \right)  G^a_{\mu\nu}~,\qquad
{\cal O}_{G2} =  \left( {\bar Q}_L \lambda_{\rm FC} \gamma_\mu T^a Q_L \right) D_\mu G^a_{\mu\nu}~.
\eeq
In principle, one could write similar structures for the electroweak gauge
bosons. However, after the spontaneous breaking  of ${\rm SU}(2)_L \otimes {\rm U}(1)_Y$,
the only terms relevant  for low-energy processes  ($p^2 \ll v^2$) are those
involving the photon field, namely
\beq
{\cal O}_{F1} =  H^\dagger \left( {\bar D}_R  \diagyuk_d \lambda_{\rm FC} \sigma_{\mu\nu}
Q_L \right) F_{\mu\nu}~,\qquad
{\cal O}_{F2} =  \left( {\bar Q}_L \lambda_{\rm FC} \gamma_\mu Q_L \right) D_\mu F_{\mu\nu}~.
\eeq

\item{$\underline{\Delta F=1\ \mbox{four-fermion operators.}}$}
The operators involving leptons are
\beqa
&& {\cal O}_{\ell 1} = \left( {\bar Q}_L \lambda_{\rm FC} \gamma_\mu Q_L \right)
        ({\bar L}_L \gamma_\mu L_L)~, \qquad
{\cal O}_{\ell 2} = \left( {\bar Q}_L \lambda_{\rm FC} \gamma_\mu  \tau^a Q_L \right)
        ({\bar L}_L \gamma_\mu  \tau^a L_L)~, \no \\
&& {\cal O}_{\ell 3} = \left( {\bar Q}_L \lambda_{\rm FC} \gamma_\mu Q_L \right)
        ({\bar E}_R \gamma_\mu E_R)~.
\eeqa
The operators involving only quarks are
\beq
\ba{ll}
{\cal O}_{q1} =  \left( {\bar Q}_L \lambda_{\rm FC} \gamma_\mu Q_L \right)
          ({\bar Q}_L  \gamma_\mu Q_L)~,
& {\cal O}_{q2} =  \left( {\bar Q}_L \lambda_{\rm FC} \gamma_\mu \tau^a Q_L \right)
         ( {\bar Q}_L  \gamma_\mu \tau^a Q_L)~, \\
{\cal O}_{q3} =  \left( {\bar Q}_L \lambda_{\rm FC} \gamma_\mu T^a Q_L \right)
         ( {\bar Q}_L  \gamma_\mu T^a Q_L)~,
& {\cal O}_{q4} =  \left( {\bar Q}_L \lambda_{\rm FC} \gamma_\mu T^a\tau^b Q_L \right)
         ( {\bar Q}_L  \gamma_\mu T^a\tau^b Q_L)~, \\
{\cal O}_{q5} =  \left( {\bar Q}_L \lambda_{\rm FC} \gamma_\mu  Q_L \right)
         ( {\bar D}_R \gamma_\mu  D_R)~,
& {\cal O}_{q6} =\left( {\bar Q}_L \lambda_{\rm FC} \gamma_\mu T^a Q_L \right)
         ( {\bar D}_R \gamma_\mu T^a D_R)~, \\
{\cal O}_{q7} =  \left({\bar Q}_L \lambda_{\rm FC} \gamma_\mu  Q_L \right)
         ( {\bar U}_R \gamma_\mu  U_R)~,
& {\cal O}_{q8} =  \left({\bar Q}_L \lambda_{\rm FC} \gamma_\mu T^a Q_L\right)
         ( {\bar U}_R \gamma_\mu T^a U_R)~.
\ea
\eeq
\end{itemize}
The number of independent $\Delta F=1$ terms is substantially reduced once
we consider the spontaneous breakdown of the gauge group,
we integrate out off-shell gauge fields, and we restrict
the attention to the down-type component of the terms between square brackets.
This corresponds to project
the ${\cal O}_i$ into the SM basis of FCNC operators:
\beq
\cH^{\Delta F=1}_{\rm eff} = \frac{1}{ \Lambda^2} \sum_{n} a_n {\cal O}_n ~+~{\rm h.c.} \qquad
\longrightarrow \qquad \frac{G_{\rm F} \alpha}{ 2\sqrt{2}\pi \sw }
 V^*_{3i} V_{3j}  \sum_{n} C_n {\cal Q}_n~+~{\rm h.c.}
\label{eq:proj}
\eeq
The sum on the r.h.s.\ of eq.~(\ref{eq:proj}) includes 13 terms 
(see appendix \ref{app:qi}),
namely four QCD-penguin operators (${\cal Q}_{3\ldots 6}$), four electroweak-penguin
operators (${\cal Q}_{7\ldots 10}$), magnetic and chromo-magnetic dipole
operators (${\cal Q}_{7\gamma}$ and ${\cal Q}_{8G}$), and three quark-lepton
operators (${\cal Q}_{9V}$, ${\cal Q}_{10A}$ and ${\cal Q}_{\nu\bar \nu}$).
In this approach, the projection (\ref{eq:proj}) defines the
leading non-standard contributions to the
initial conditions of the $C_i$ at the electroweak scale.
Note that the normalization of the r.h.s.\  of (\ref{eq:proj})
is such that the pure electroweak contribution to
the $C_i(M^2_W)$ is of order one.
Since we assumed that CP is broken only by the background values of $\yuk$,
the $a_i$ coefficients are real.

Defining
\beq
\epsilon_i = \left( \frac{\Lambda_0}{\Lambda} \right)^2 a_i~, \qquad
\Lambda_0 = \frac{\diagyuk_t \sw  M_W}{\alpha } \approx  2.4 ~{\rm TeV}~,
\label{eq:eps_i}
\eeq
in the case of quark-lepton and electroweak-penguin operators
we find
\beqa
\delta C_{\nu\bar\nu}
&=& \epsilon_Z + \epsilon_{\ell 1} -\epsilon_{\ell 2}  \label{eq:Cnueps} \\
\delta C_{10A} &=& \epsilon_Z -\epsilon_{\ell 1} -\epsilon_{\ell 2} +\epsilon_{\ell 3} \\
\delta C_{9V} &=& \epsilon_{\ell 1}+\epsilon_{\ell 2}+\epsilon_{\ell 3} -
\left[ (1-4\sw) \epsilon_Z + 2 e \epsilon_{F2} \right] \label{eq:C9V}  \\
\delta C_7  &=&  - \epsilon_{q5} + \epsilon_{q7}
+ \frac{1}{2N_c} (\epsilon_{q6}-\epsilon_{q8}) -\frac{1}{2} \epsilon_Z + \frac{1}{2}
\left[ (1-4\sw)\epsilon_Z + 2 e\epsilon_{F2}\right]  \\
\delta C_8  &=& -\frac{1}{2}(\epsilon_{q6}-\epsilon_{q8})  \\
\delta C_9  &=&   -2 \epsilon_{q2}  + \frac{1}{N_c} \epsilon_{q4} + \frac{3}{2}\epsilon_Z
+ \frac{1}{2}\left[ (1-4\sw)\epsilon_Z  +  2 e \epsilon_{F2} \right]  \\
\delta C_{10}  &=&   -\epsilon_{q4}  \label{eq:C10eps}
\eeqa
where  $\delta C_i = C_i(M_W^2)- C_i^{\rm SM} (M_W^2)$ and
$\epsilon_Z = (\epsilon_{H1}+\epsilon_{H2})/2$. Some comments are in order:
\begin{itemize}
\item{}
The seven combinations of $\epsilon_i$ appearing in
eqs.~(\ref{eq:Cnueps}--\ref{eq:C10eps}) are all
independent (even if the suppressed term between square brackets
is neglected). This result was not obvious a priori, since the
${\cal Q}_i$ basis is not ${\rm SU}(2)_L \otimes {\rm U}(1)_Y$ invariant.
\item{}
The leading SM contribution to the above $C_i$, at the electroweak
scale, can be expressed as
\beqa
&& \epsilon^{\rm SM}_{\ell 2} = \epsilon^{\rm SM}_{q2} = Z_0(x_t)
      - \left[ X_0(x_t)+Y_0(x_t)\right]/2~, \qquad
   \epsilon^{\rm SM}_{\ell 1} = \left[X_0(x_t)-Y_0(x_t)\right]/2~, \no \\
&& \epsilon^{\rm SM}_Z = Z_0(x_t)~, \qquad
   \epsilon^{\rm SM}_{F2} = \epsilon^{\rm SM}_{\ell 3} =
   \epsilon^{\rm SM}_{q3 \ldots q8} =0~, \no
\eeqa
where $X$, $Y$ and $Z$ are the $\cO(1)$ loop functions
defined as in~\cite{BBL}. Thus, in general, the non-standard contributions
cannot be re-absorbed into a redefinition of the SM electroweak
contributions.
\item{}
Since the $\epsilon_i$ appear with $\cO(1)$ coefficients in
eqs.~(\ref{eq:Cnueps}--\ref{eq:C10eps}), an experimental determination
of the parameters $C_i$, at the weak scale, with a precision $p$, allow,
in general, to set bounds of $\cO(\Lambda_0/\sqrt{p})$ on the effective
scale of dimension-six operators.
Thus precision experiments on rare decays
could aim to probe effective scales of new physics up to $\sim 10$~TeV,
within MFV models.\footnote{~When 
comparing with apparently stronger bounds in the literature, 
one should notice that we assume $1/\Lambda^2$  
as coefficient of the operators, rather than $4\pi/\Lambda^2$.}
\item{}
The seven independent combinations appearing in
eqs.~(\ref{eq:Cnueps}--\ref{eq:C10eps}) could
be determined by experimental data on one type
of $d_i \to d_j$ amplitudes only. Then the
MFV hypothesis could be tested
by comparing different types of FCNC transitions
(namely $b \to d$, $b \to s$ and  $s \to d$).
\end{itemize}
In the case of $C_{3\ldots 6}$ the SM contribution is
strongly enhanced by QCD and is likely to obscure
any effect of the $\epsilon_i$, if these are $\cO(1)$ or smaller.
On the other hand, the  $\epsilon_i$ have a potential non-negligible
impact on the initial conditions of dipole operators,
\beq
\delta C_{7\gamma} = \frac{2 g^2}{e} \epsilon_{F1}~, \qquad \qquad
\delta C_{8G} = \frac{ 2 g^2}{g_s} \epsilon_{G1}~,
\label{eq:C7eps}
\eeq
whose SM contributions are
\beqa
C^{\rm SM}_{7\gamma}(M_W^2) = - \frac{1 }{2 }D'_0(x_t) ,
\qquad \qquad C^{\rm SM}_{8G}(M_W^2) = - \frac{1}{2} E'_0(x_t)~. \no
\eeqa
The loop functions $D'_0$ and $E'_0$ are given in~\cite{BBL}.
The last term needed for a complete analysis of the flavour sector
is the universal modification of $\Delta F=2$ amplitudes.
This leads to
\beq
\delta C_0 = 2 \epsilon_0~,
\label{eq:C0eps}
\eeq
where
\beq
\cH^{\Delta F=2}_{\rm eff} = \frac{G^2_F M_W^2}{ 8 \pi^2 }
 (V^*_{3i} V_{3j})^2  C_0 \left[ \bar d_i \gamma_\mu (1-\gamma_5) d_j \right]^2~,
\eeq
so that $C_0^{\rm SM}(M_W^2) = S_0(x_t)/2$~\cite{BBL}.

In conclusion, the leading non-standard FCNC effects in (one-Higgs doublet)
models with MFV
can be described in terms of the ten independent
$\epsilon_i$ combinations appearing in eqs.~(\ref{eq:Cnueps})--(\ref{eq:C0eps}).

\section{Experimental bounds}
\label{sect:bounds}

\renewcommand\arraystretch{1.2}
\begin{table}
$$
\begin{array}{rlc|ccc}
\multicolumn{2}{c}{\hbox{Minimally flavour violating}}\Black &\hbox{main}
&\multicolumn{2}{c}{\Lambda\hbox{ [TeV]}}&\\
\multicolumn{2}{c}{\hbox{dimension~six~operator}}\Black &
\hbox{observables}\Black &
-\Black& +\Black\\
\hline
\cO_{0}= &\frac{1}{2} (\bar Q_L  \lambda_{\rm FC} \gamma_{\mu} Q_L)^2 \Black
\phantom{X^{X^X}}
&\epsilon_K,\quad \Delta m_{B_d}      &6.4 & 5.0 \\
\cO_{F1}= &   H^\dagger \left( {\bar D}_R  \diagyuk_d \lambda_{\rm FC} \sigma_{\mu\nu}
Q_L \right) F_{\mu\nu}  \Black &
B\to X_s \gamma     &9.3 & 12.4 \\
\cO_{G1}= &  H^\dagger \left( {\bar D}_R  \diagyuk_d \lambda_{\rm FC} \sigma_{\mu\nu}
  T^a  Q_L \right)  G^a_{\mu\nu}  \Black &
B\to X_s \gamma & 2.6 & 3.5 \\
\cO_{\ell1}=& (\bar Q_L  \lambda_{\rm FC}\gamma_{\mu}   Q_L)(\bar L_L \gamma_\mu L_L )  \Black
&B\to (X) \ell\bar{\ell},\quad K\to \pi \nu\bar{\nu},(\pi) \ell \bar{\ell} \quad
  & 3.1 & 2.7  & *\\
\cO_{\ell2}= &( {\bar Q}_L \lambda_{\rm FC} \gamma_\mu  \tau^a Q_L)
 ({\bar L}_L \gamma_\mu  \tau^a L_L)\quad   \Black &B\to (X) \ell\bar{\ell},\quad
 K\to \pi \nu\bar{\nu},(\pi) \ell \bar{\ell} \quad  & 3.4 & 3.0  & *\\
\cO_{H1}=& (\bar Q_L  \lambda_{\rm FC} \gamma_{\mu} Q_L)(H^\dagger i D_\mu H)\qquad  \Black
&B\to(X) \ell\bar{\ell},\quad K\to \pi \nu\bar{\nu},(\pi) \ell \bar{\ell} \quad
 &  1.6 & 1.6 & *\\
\cO_{q5}=& (\bar Q_L  \lambda_{\rm FC} \gamma_{\mu} Q_L)(\bar D_R \gamma_\mu D_R )  \Black
&B\to K\pi,\quad \epsilon'/\epsilon,\ldots   &    \multicolumn{2}{c}{\sim 1} \\
\end{array}$$
\caption[X]{\label{tab:tab}\em
{\bf Bounds on MFV operators}.
The SM is extended by adding minimally flavour-violating
dimension-six operators with coefficient $\pm1/\Lambda^2$
($+$ or $-$ denote their constructive or destructive
interference with the SM amplitude). Here
we report the bounds at $99\%$ {\rm CL} on $\Lambda$, in $\TeV$,
for the single operator (in the most representative cases).
Fine-tuned scenarios with small $\Lambda$, such that new physics
flips the sign of the SM amplitude are not reported (see text).
The $*$ signals the cases where a significant
increase in sensitivity is expected in the near future.
}
\end{table}
\renewcommand\arraystretch{1}

\begin{figure}[t]
$$
\includegraphics[width=7.3cm]{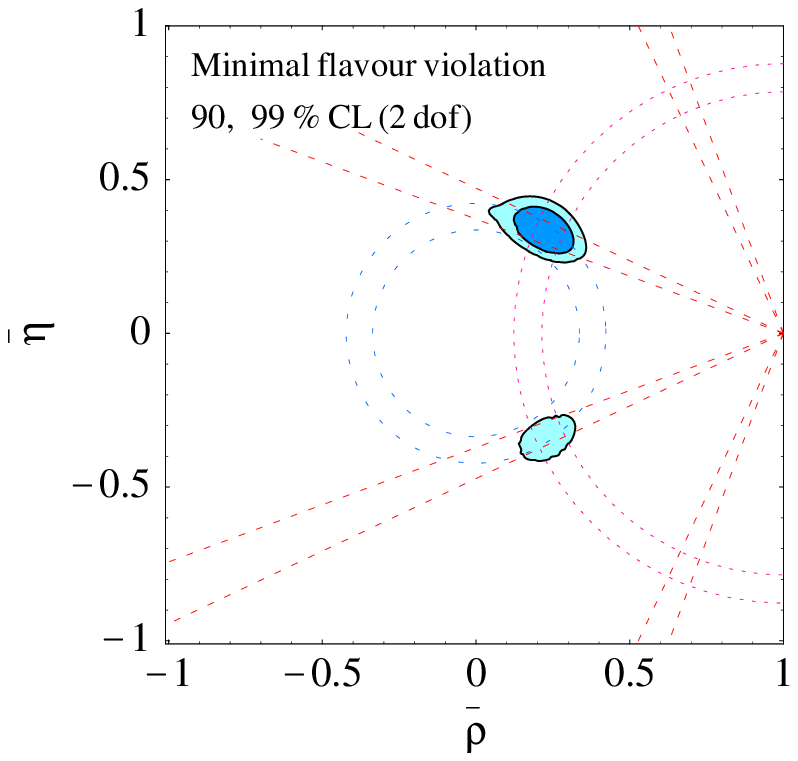}\hspace{1cm}
\includegraphics[width=7cm]{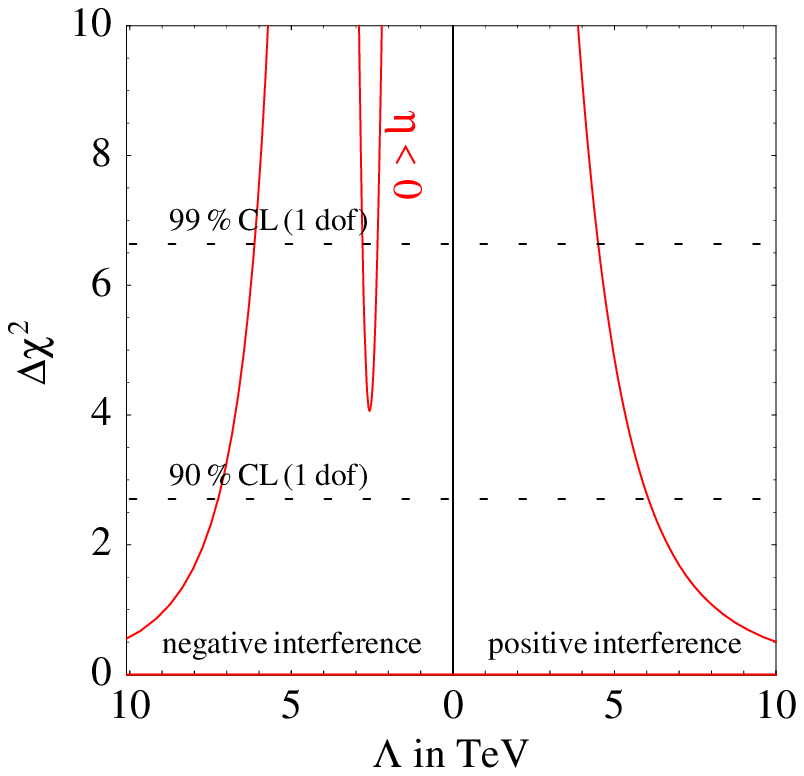}
$$
\caption[X]{\em {\bf Fit of $\Delta F=2$ data in MFV models}.
Left: fit of $\bar \rho$ and $\bar \eta$ in models with minimal
flavour violation. The dotted lines denotes 68\% CL intervals
imposed by the observables insensitive to $|C_0(M_W^2)|$,
namely $|V_{ub}|$,  $\Delta M_{B_d}/\Delta M_{B_s}$ and $a^{CP}_{\Psi K}$.
Input values are shown in table~\ref{tab:data}.
Right: $\Delta \chi^2$ of the global CKM fit as a function of the
scale of the operator ${\cal O}_0$.
\label{fig:CKMfit}}
\end{figure}

\begin{table}[t]
$$\begin{array}{rcll}
m_t(m_t)  &=& (166.8\pm5.1)\GeV &                 \hbox{running top masss}\\
m_b(m_b)  &=& (4.23\pm 0.07)\GeV &                 \hbox{running bottom masss}\\
m_c(m_c)  &=& (1.3\pm0.1)\GeV &                   \hbox{running charm masss}\\
\alpha_{\rm s}(M_Z)  &=& 0.119\pm0.003 &          \hbox{strong coupling}\\
\Delta M_{B_d} &=& (0.487\pm0.014)/\hbox{ps} &  \hbox{$B_d$ mass difference}\\
\lambda &=& 0.2237\pm0.0033&                     \hbox{Cabibbo angle}\\
A &=& 0.819\pm0.035&                             \hbox{Wolfenstein $A$ parameter}\\
\left|V_{ub}/V_{cb}\right| &=& 0.0865\pm 0.0094 &  \hbox{ratio of CKM mixings}\\
a^{CP}_{\Psi K} &=& 0.732\pm0.063   
   &  \hbox{CP-violation in $B_d\to \Psi K$ \protect\cite{sin2b}}\\
\hline
f_{B_d} B_{B_d}^{1/2} &=& (0.230\pm 0.025)\GeV &                          \hbox{Matrix element for $B_d$ mixing} \\
\xi &=&  1.14\pm0.04&                             \hbox{Relative matrix element for $B_s$ vs $B_d$ mixing}  \\
B_K &=& 0.87 \pm 0.06 &                           \hbox{matrix element for $K$ mixing} \\
\eta_1 &=& 1.38 \pm 0.25 &                           \hbox{RGE factor for $cc$-induced $K$ mixing} \\
\eta_2 &=& 0.57 \pm 0.01 &                           \hbox{RGE factor for $tt$-induced $K$ mixing} \\
\eta_3 &=& 0.47 \pm 0.04 &                           \hbox{RGE factor for $ct$-induced $K$ mixing} \\
\eta_B &=& 0.55 \pm 0.01 &                           \hbox{RGE factor for $B$ mixing} \\
\end{array}$$
\caption{\em Data used to produce the CKM fit
in fig.~\ref{fig:CKMfit}. The probability distribution of
$\Delta M_{B_s}$, as well as input values not reported
above, are taken from~\cite{CKMfit}.
\label{tab:data}
}
\end{table}

\subsection{$\Delta F=2$ and the CKM fit}
In generic models with MFV, the CKM matrix
can still be determined with good accuracy~\cite{UUT0,UUT2}.
Employing the improved Wolfenstein
parametrization~\cite{Buras_CKM},
$\lambda$ and $A$ are completely determined by tree-level
processes and only  $\bar \rho$ and $\bar \eta$ are
potentially affected by new physics. The data used to
constrain $\bar \rho$ and $\bar \eta$ are
$|V_{ub}|$, determined from a tree-level
transition, $\Delta M_{B_d}/\Delta M_{B_s}$, $a^{CP}_{J/\Psi K}$,
$\Delta M_{B_d}$ and $\epsilon_K$, which are sensitive
to $\Delta F=2$ amplitudes. If the
modification of $\Delta F=2$ amplitudes
is induced only by the operator ${\cal O}_0$ in eq.~(\ref{eq:O0}),
then the ratio  $\Delta M_{B_d}/\Delta M_{B_s}$ remains unchanged
and the relation between $a^{CP}_{\Psi K}$ and
$\sin(2\beta)$ can only be modified by an
overall sign~\cite{UUT2}.

In fig.~\ref{fig:CKMfit} (left) we show
the result of a $\bar \rho$--$\bar \eta$ fit
where $C_0(M_W^2)$, or the Wilson coefficient of ${\cal O}_0$
at the electroweak scale, is treated as a free parameter.
As can be noted, in addition to the standard solution
with $\bar \eta>0$, also a solution with $\bar \eta<0$
appears. The latter arises from the fine-tuned scenario where
$C_0(M_W^2)$  is about equal in magnitude to the SM case
but has the opposite sign.
The solution with $\bar \eta<0$ is somewhat disfavoured,
as can be more precisely seen in the right panel of fig.~\ref{fig:CKMfit}.
In fact, while the SM contribution to $\epsilon_K$ proportional to the charm mass
cannot be neglected because is enhanced by infra-red logarithms,
the new physics contribution to $\epsilon_K$ is dominated by the top, see eq.~(\ref{eq:FC}).
This breaks the $(C_0, \bar\eta)
\leftrightarrow (-C_0,-\bar\eta)$ symmetry of all
 other observables.

\medskip

The low-quality solution with
$\bar \eta <0 $ is obtained for  $\Lambda \approx \Lambda_0$.
Barring this fine-tuned possibility,
the results in fig.~\ref{fig:CKMfit} can be translated
into a bound on the effective scale of the non-standard
$\Delta F=2$ operator,
reported in the first row of table~\ref{tab:tab}.
It will be difficult to improve these bounds in the near future,
because hadronic
uncertainties on the matrix elements of $\Delta S=2$ and
$\Delta B=2$ operators are the main limiting 
factor.\footnote{~In the global fit of fig.~\ref{fig:CKMfit} we 
use only the statistical error quoted in ref.~\cite{CKMfit} 
for the hadronic parameters $f_B$ and $B_{B,K}$
measured on the lattice (see table \ref{tab:data}).
Doubling the errors, to take into account possible systematic effects, 
would not qualitatively modify the fit and, in particular, 
would weaken the bounds on the scale $\Lambda$
of ${\cal O}_0$ at most by $30\%$.}
Even with $\cO(1\%)$ experimental uncertainties on
$a^{CP}_{\Psi K}$ and $\Delta M_{B_d}/\Delta M_{B_s}$,
the $\Delta F=2$ bounds on $\Lambda$ will remain below
10 TeV as long as the theoretical errors on the
matrix elements will remain above $10\%$.
The constraint from $|V_{ub}/V_{cb}|$ (circle centered on the origin)
around the best fit regions is consistent and tangent to the constraint from
$a^{CP}_{\Psi K}$ (rays originating from $\bar\rho=1$), and therefore has little impact in the global fit.

\subsection{$B \to X_s \gamma$}
\begin{figure}[t]
$$
\includegraphics[width=7cm]{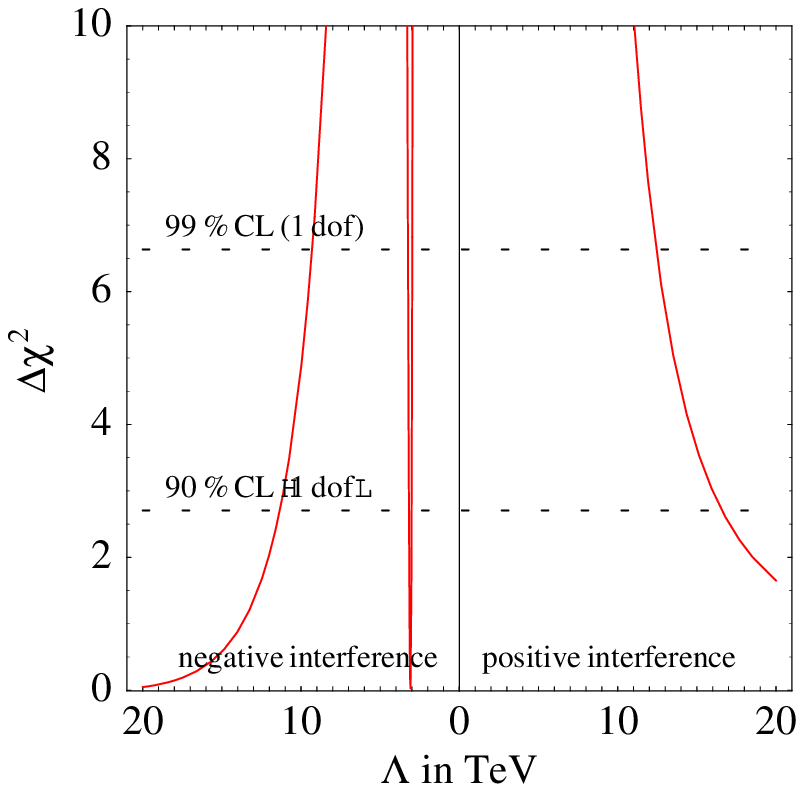}\hspace{1cm}
\includegraphics[width=7cm]{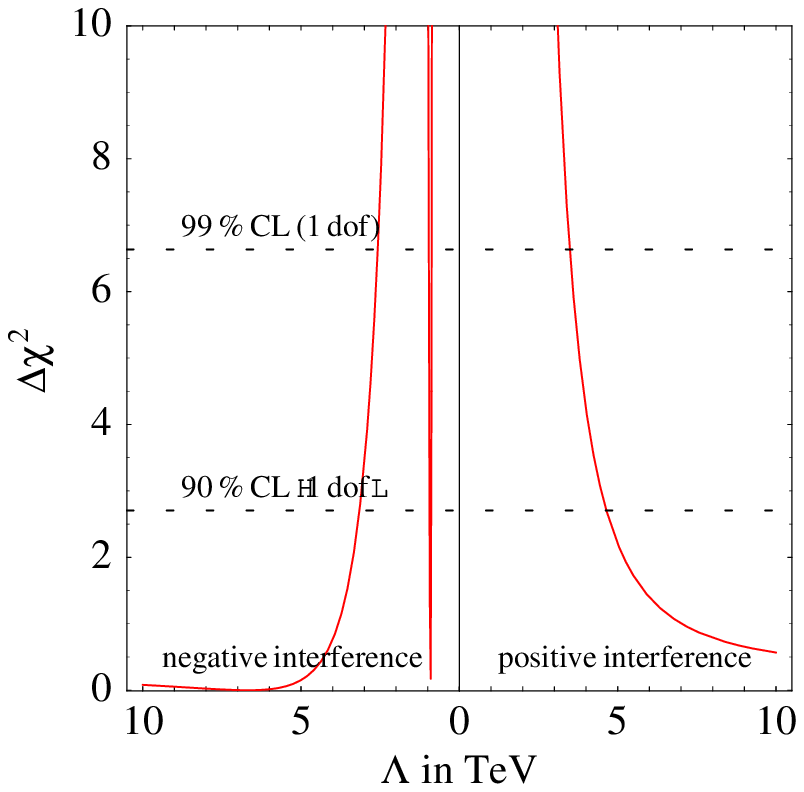}
$$
\caption[X]{\em  Bounds on the
scale of the operators $\cO_{F1}$ (left) and
$\cO_{G1}$ (right) from $B\to X_s \gamma$.
\label{fig:bsgfit}}
\end{figure}

The inclusive rare decay $B \to X_s \gamma$ provides, at present, the most
stringent bound on the effective scale of the new FCNC operators.
Taking into account the new precise measurements reported
by CLEO~\cite{CLEO_bsg}, Belle~\cite{Belle_bsg}, and Babar~\cite{Babar_bsg}
the world average is
\beq
\BR ( B \to X_s \gamma ) = 
 (3.46 \pm 0.34) \times 10^{-4}~,
\label{eq:bsg_exp}
\eeq
to be compared with the SM expectation~\cite{MisiakG}\footnote{~We adopt 
the theoretical error of ref.~\cite{MisiakG}, which we consider 
as a reasonable estimate. Doubling the theoretical 
error would weaken the bounds on the scale $\Lambda$
that suppresses the $b\to s\gamma$ operators
by about $20\%$.}
\beq
\BR (B \to X_s \gamma)_{\rm SM} =  (3.73 \pm 0.30) \times 10^{-4}~.
\label{eq:bsg_MG}
\eeq
As is well known, NLO QCD corrections play a rather
important role in this process. For this reason, we will
assume that $\delta C_{7\gamma,8G}$,
defined in eq.~(\ref{eq:eps_i}), describe the deviations of
$C_{7\gamma,8G}^{\rm NLO}(M_W^2)$ from their SM values,
although this is formally not correct since QCD
corrections to the new effective operators have
not been included.
The full NLO expression for $\BR (B \to X_s \gamma)$ 
(see refs.~\cite{MisiakG,bsg_NLO,CDGG}),
used in our numerical computation, 
can be approximated as
\beq
\frac{ \BR (B \to X_s \gamma)}{ \BR (B \to X_s \gamma)_{\rm SM} }
 =  1 - 2.4\delta C_{7\gamma} - 0.7\delta C_{8G} + {\cal O}(\delta C^2)~.
\label{eq:bsg_new}
\eeq
Since the error in eq.~(\ref{eq:bsg_MG}) is largely dominated by
low-energy dynamics (in particular by the value of $m_c$), we ignore
the correlations between $\BR (B \to X_s \gamma)_{\rm SM}$
and the coefficients on the r.h.s.\ of eq.~(\ref{eq:bsg_new}).
Combining all errors in quadrature and setting, alternatively,
$\delta C_{8G}=0$ or $\delta C_{7\gamma}=0$, we obtain the fits
shown in fig.~\ref{fig:bsgfit} and the bounds reported in table~\ref{tab:tab}.
Also in this case there is a particular value of $\Lambda$ for which we
obtain the condition
$C_{7\gamma}(m_b)\approx -C_{7\gamma}^{\rm SM}(m_b)$,
allowed by data. This corresponds to the sharp peaks shown
in fig.~\ref{fig:bsgfit}, located at $\Lambda \approx 3$~TeV for
$\cO_{F1}$ and $\Lambda \approx 1$~TeV for
$\cO_{G1}$.

In presence of both $\delta C_{8G}\not=0$ and $\delta C_{7\gamma}\not=0$
there can be cancellations between their effects,
giving rise to two allowed bands for their coefficients.
When studying processes related to $b\to s \gamma$,
such as  $b\to s \bar\ell \ell$, one has to keep in mind
that the effective constraint is that the renormalized coefficient $|C_{7\gamma}(m_b)|$
must be very close to the its SM value.

\medskip

It is interesting to compare the limits on the effective scale of ${\cal O}_{F1}$
with limits on flavour-diagonal dipole-type operators.
Since we assumed no new sources of CP violation, we get no useful bounds from
electric-dipole moments. On the other hand, the anomalous magnetic moment
of the muon does provide a bound on the operator
$e H^\dagger 
\left(\bar E_R  \diagyuk_\ell \sigma_{\mu\nu} L_L \right) F_{\mu\nu}$.
Taking into account theoretical and experimental uncertainties,
the limit on its effective scale is around 2 TeV, much
weaker than in the  $B \to X_{s}\gamma$  case.

We finally note that in this approach
the stringent constraints from $B \to X_s \gamma$ can
directly be translated into bounds on non-standard effects
in $s \to d \gamma$ and $b \to d \gamma$ transitions.
The former could in principle tested by CP-asymmetries in
$K \to \pi\pi\gamma$ decays~\cite{DI}. However,
due long-distance contaminations, only an order-of-magnitude
enhancement of the short-distance amplitude
could eventually be seen in $K \to \pi\pi\gamma$: the constraint from
$B \to X_s \gamma$ implies this cannot occur in MFV models.
More promising from the experimental point of view
is the inclusive $B \to X_{d}\gamma$ rate, whose
bound in MFV models is shown in table~\ref{tab:neg_lim}.

\renewcommand\arraystretch{1.2}
\begin{table}
$$
\begin{array}{l|lc|l}
\hbox{Observable }  & \hbox{99(90)\%  MFV limit}\!\!\!\!\!\! 
  & \ \ \hbox{determined by} & \hbox{90\% exp. limit ~\cite{PDG}}   \\ \hline
\BR(B \to X_{d}\gamma) & 2(2)\times 10^{-5}  & \BR(B \to X_{s}\gamma)
  & - \\ \hline
\BR(B\to X_s \nu\bar{\nu}) & 8(6) \times 10^{-4}
  &   & 6.4 \times 10^{-4}  \\
\BR(B\to X_d \nu\bar{\nu}) & 3(2) \times 10^{-5}
  & \BR(K^+ \to \pi^+ \nu\bar{\nu})  &  -  \\
\BR(K_L \to \pi^0 \nu\bar{\nu})  &  5(4) \times 10^{-10}
  &  &  5.9 \times 10^{-7} \\
 \hline
\BR(B_s \to \tau^+ \tau^- ) & 4(3)\times 10^{-6}
  &   & -  \\
\BR(B_d \to \tau^+ \tau^-) & 1(1) \times 10^{-7}
  & \BR(B \to X_s \bar\ell \ell)   & - \\
\BR(B_s \to \mu^+ \mu^-) &  2(1)  \times 10^{-8}
  & [\BR(K_L \to \mu^+\mu^-)_{\rm short}]~  &  2.0 \times 10^{-6} \\
\BR(B_d \to \mu^+ \mu^-) &  6(5) \times 10^{-10}
  &   &  2.0 \times 10^{-7} \quad
\cite{Babar_mm} \\
\BR(K_L \to \pi^0 e^+e^-)_{\rm CP-dir}\  & 2(2) \times 10^{-11}
  &    &  5.6 \times 10^{-10} \\
\end{array}
$$
\caption[X]{\label{tab:neg_lim}\em Bounds on various
decay rates obtained in MFV models with one Higgs doublet.}
\end{table}
\renewcommand\arraystretch{1}

\subsection{Rare FCNC decays into a lepton pair}
These processes provide constraints
on the three Wilson coefficients $C_{\nu\bar\nu}$,
$C_{9V}$ and $C_{10A}$ which, in turn, receive
non-standard contributions from
${\cal O}_{H1,H2}$, ${\cal O}_{F1,F2}$ and ${\cal O}_{\ell 1\ldots \ell 3}$,
as shown in eqs.~(\ref{eq:Cnueps})--(\ref{eq:C9V}).
The simplest cases are the decays into a neutrino pair,
sensitive only to $C_{\nu\bar\nu}$, or the helicity
suppressed pure-leptonic decays [$K_L(B) \to \ell^+\ell^-$],
sensitive only to $C_{10A}$.

\begin{figure}[t]
$$
\includegraphics[width=14cm]{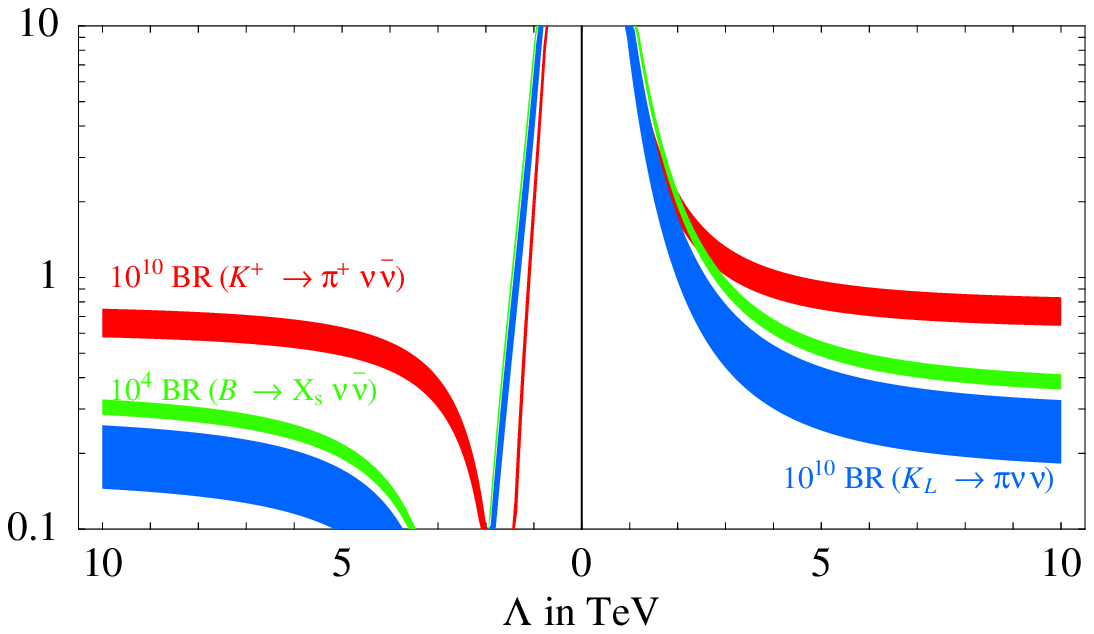}
$$
\caption[X]{\label{fig:Kpnn}\em Branching ratios of 
various processes involving $d_i\to d_j \nu\bar
\nu$ transitions as functions of the scale of the effective operator
$\cO_{\ell 1}$. The bands represent $1\sigma$ uncertainties, taking into
account the present determination of the CKM parameters.}
\end{figure}

\begin{figure}[t]
$$
\hskip -0.8 cm
\includegraphics[width=10.0cm,height=6.5cm]{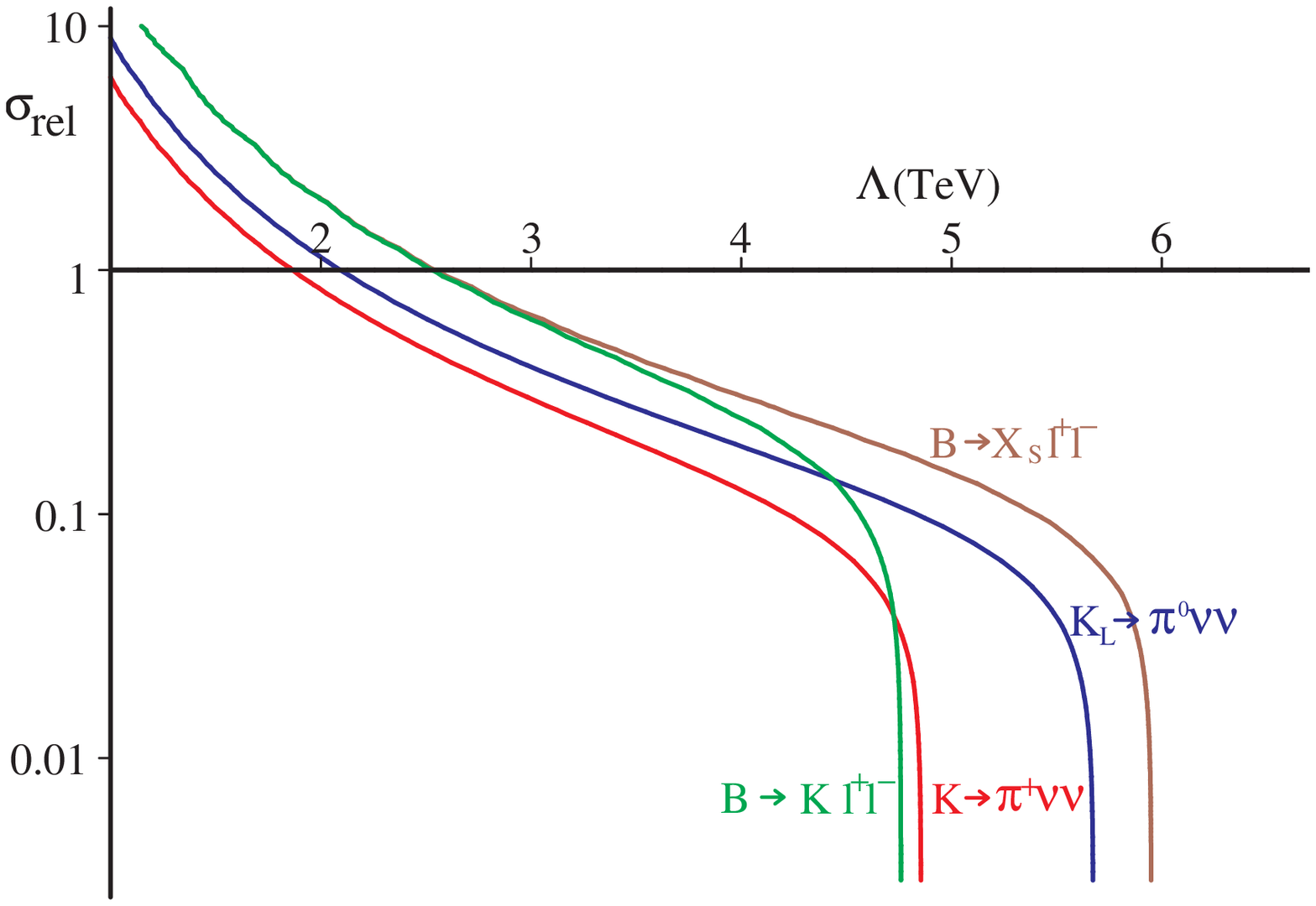}
\hskip -1.6 cm
\includegraphics[width=10.0cm,height=6.5cm]{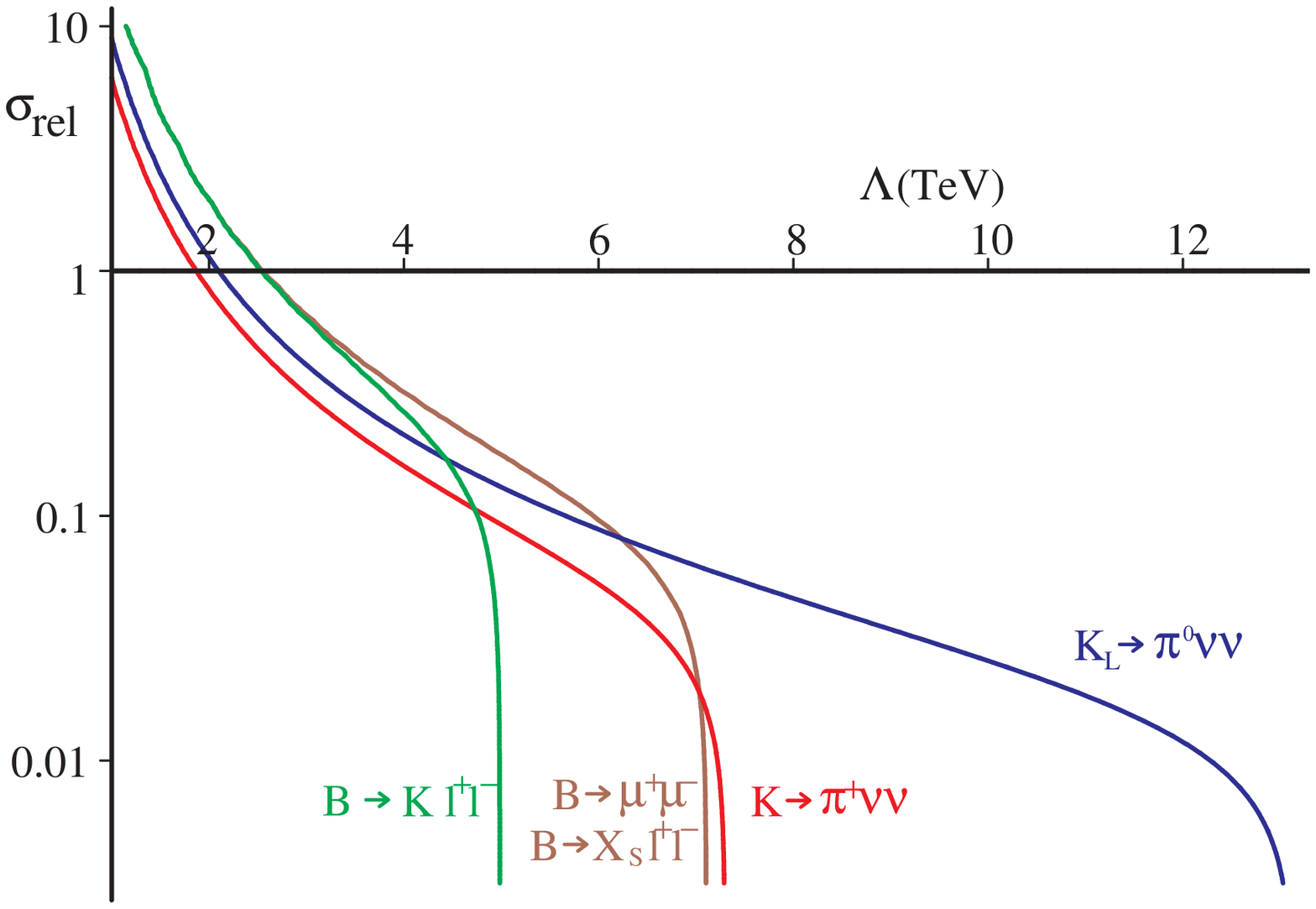}
\hskip -1.2 cm
$$
\vskip -1 cm
\caption[X]{\label{fig:conf}\em Comparison of the effectiveness 
of different rare modes
in setting future bounds on the scale of the operator $\cO_{\ell 1}$.
The vertical axis indicates the relative precision of an
hypothetic measurement of the rate, with central value equal to
the SM expectation. The curves in the two panels are obtained
assuming an uncertainty of 10\% (left) or 1\% (right) on the
corresponding overall CKM factor (e.g. $|V_{tb}^* V_{td}|^2$
in the case of $B_d\to\mu^+\mu^-$).
}
\end{figure}

\medskip

Concerning $\nu\bar\nu$ modes, the
most significant bounds come from $\BR(K^+\to\pi^+ \nu\bar{\nu})$.
Using the SM expression of this observable~\cite{BB2}
and treating $C_{\nu\bar\nu}$ as a free parameter,
the recent E787 result~\cite{E787}
\beq
\BR(K^+\to\pi^+ \nu\bar{\nu}) = \left( 1.57^{~+~1.75}_{~-~0.82} \right) \times 10^{-10}~.
\label{eq:E787}
\eeq
leads to the $99\%$ CL limit
\beq
-4.8<C_{\nu\bar\nu}/C^{\rm SM}_{\nu\bar\nu}<4.0~.
\eeq
These bounds are still rather weak, corresponding to effective scales
of the new physics operator around 1 TeV, nonetheless they
already imply significant bounds on $\BR(K_L \to \pi^0 \nu\bar{\nu})$ and
$\BR(B\to X_s \nu\bar{\nu})$, as shown in table~\ref{tab:neg_lim}
(see also ref.~\cite{Bergmann}).
In few years, at the end of the E949 experiment, assuming that
the central value
of $\BR(K^+\to\pi^+ \nu\bar{\nu})$ will move towards the SM prediction,
we can expect to probe $\Lambda$ from this observable
up to about $3$ TeV, a sensitivity comparable to the one of
electroweak precision data on the corresponding
flavour-diagonal terms. At the end of the CKM experiment,
with a $10\%$ measurement of this branching ratio,
the search on $\Lambda$ could be pursued up to above 5 TeV.
As illustrated in fig.~\ref{fig:conf}, a
sensitivity 
above 10 TeV could in principle
be reached by a measurement of
$\BR(K_L \to\pi^0 \nu\bar{\nu})$ with a precision of few percent around
the SM value, since this rate
is free from the theoretical uncertainty
due to charm contributions.

\medskip

The most interesting constraint on pure-leptonic decays is
provided by $K_L \to \mu^+\mu^-$. The branching ratio
of this decay is measured very precisely;
however, this mode receives also a large long-distance amplitude
by the two-photon intermediate state. Actually short- and long-distance
dispersive parts cancel each other to a good extent, since the total
$K_L\to \mu^+\mu^-$ rate is almost saturated by the absorptive
two-photon contribution. Taking into account the recent experimental
result on  $\BR(K_L \to \mu^+\mu^-)$~\cite{E871} and following
the analyses of the long-distance amplitude in~\cite{DIP},
we shall impose the conservative upper bound
\beq
\BR(K_L\to\mu^+\mu^-)^{\rm short} \le 3.0 \times 10^{-9}~,
\eeq
that we shall treat as an absolute limit. Employing the SM expression of
$\BR(K_L\to\mu^+\mu^-)^{\rm short}$ \cite{BBL} and taking into account the uncertainty
of the CKM fit, this information can be translated into the $99\%$ CL limit
\beq
-3.2 <C_{10A}/C^{\rm SM}_{10A}< 2.6~.
\label{eq:C10Kmm}
\eeq
We then conclude that the present level of precision reached by
$K_L \to \mu^+\mu^-$ and $K^+\to\pi^+\nu\bar\nu$ is comparable,
although the bounds from the former process are still more stringent.
However, it should stressed that in the $K_L \to \mu^+\mu^-$ case
the limiting error is theoretical and very difficult to substantially
improve in the near future.

\medskip

Differential rate and forward-backward asymmetries in
$B \to X_s \ell^+ \ell^-$ offer, in principle, the possibility to
disentangle the contributions of both $C_{9V}$ and $C_{10A}$,
as well as their relative sign with $C_{7\gamma}$~\cite{AGM}.
Experiments are still far from such a detailed analysis,
nonetheless recent data from Belle~\cite{Belle_rare} and
Babar~\cite{Babar_rare} already start to provide significant
constraints on these coefficients.
Referring to~\cite{Ali01} for a detailed discussion of the
interplay of the various measurements, here we shall limit
to analyse the consequences of the most interesting result, namely
\beq
\BR (B \to K \ell^{+} \ell^{-})
 = (0.76 \pm 0.19)\times 10^{-6}~\cite{Belle_rare,Babar_rare}~.
\label{eq:BKll_lim}
\eeq
Employing the $B\to K$ form factors in~\cite{Ali99},
the $B \to K \ell^{+} \ell^{-}$ rate can be fitted using the
following approximate expression
\beq
\BR (B \to K \ell^+ \ell^-)= \left[ 2.8 ( 1.0-\delta C_{10A} )^2 +
 2.7 ( 1.1 + \delta C_{9V} \pm 0.13 )^2 \right] \times 10^{-7}~,
\label{eq:cvca}
\eeq
where the overall normalization is affected by a $30\%$ theoretical uncertainty.
The $\pm$ term in eq.~(\ref{eq:cvca}) corresponds to the small contribution
of $C_7(m_b)$, after the $B\to X_s\gamma$ constraint has been imposed
(the negative sign corresponds to the SM solution). Setting to zero the
vector contribution (i.e. in the most conservative scenario)
we find a $99\%$~CL bound on $C_{10A}$ slightly more stringent
than in eq.~(\ref{eq:C10Kmm}):
\beq
-2.3 <C_{10A}/C^{\rm SM}_{10A}< 2.3~.
\label{eq:C10BK}
\eeq
Given the important role of hadronic uncertainties in the derivation of
both eq.~(\ref{eq:C10BK}) and eq.~(\ref{eq:C10Kmm}), we do not
combine these two constraints on $C_{10A}$: the MFV bounds on
$B\to \ell^+\ell^-$ reported in table~\ref{tab:neg_lim}
are obtained from eq.~(\ref{eq:C10BK}) only.
Using eqs.~(\ref{eq:BKll_lim})--(\ref{eq:cvca}) we also derived
a rather stringent bound on the direct-CP-violating contribution to
$K_L \to \pi^0 e^+ e^-$, sensitive both
to $C_{9V}$ and $C_{10A}$~\cite{BBL}.

\medskip

The constraints on $C_{\nu\bar\nu}$, $C_{10A}$ and $C_{9V}$
can finally be combined to derive significant bounds on the
effective scale of the dimension-six operators
${\cal O}_{H1,H2}$ and ${\cal O}_{\ell 1 \ldots \ell 3}$.
We report in table~\ref{tab:tab} the results of
three representative cases: $O_{H1}$, which
modifies in the same way $C_{\nu\bar\nu}$ and $C_{10A}$,
leaving $C_{9V}$ almost unaffected;  $O_{\ell 2}$, which
increases or decreases simultaneously the three
coefficients; $O_{\ell 1}$, which acts constructively
on $C_{10A,9V}$ and destructively on $C_{\nu\bar\nu}$,
or viceversa (the first possibility is reported under the +
sign in table~\ref{tab:tab}).

Focusing on $d_i \to d_j\nu\bar \nu$ transitions,
we plot in fig.~\fig{Kpnn}  the predicted rates for
the  $K^+\to\pi^+\nu\bar\nu$, $K_L\to \pi^0 \nu\bar\nu$ and 
$B\to X_s \nu\bar{\nu}$
decay rates in presence of the ${\cal O}_{\ell1}$ operator.
The left and right panels correspond to destructive and constructive 
interference with the SM contribution.
The bands show the present one standard deviation uncertainty,
combining all errors in quadrature.
Improved determinations of the CKM parameters will make
the predictions more precise. Since these modes are all affected
by the same combination of effective operators, the presence 
of other operators, such as $\cO_{H 1}$ or $\cO_{\ell 2}$, 
induces only a rescaling of the horizontal axis.

In view of future precise measurements of rare modes,
in fig.~\ref{fig:conf} we compare the effectiveness 
of different channels in setting bounds on the scale 
of ${\cal O}_{\ell1}$. This figure should be taken with 
some care since the bounds strongly depend on the 
central values of the measurements (here assumed to 
coincide with SM expectations) and the comparison 
is modified in presence of additional operators. 
However, some interesting conclusions can still be drawn. 
For instance, it is clear that in the short term
an important r\^ole will be played by $B\to X_s \ell^+ \ell^-$: 
a $10\%$ measurement of this rate --- within the reach of $B$
factories --- should lead to bounds on $\Lambda$ up to above 5 TeV.
On the other hand, $K_L \to\pi^0 \nu\bar{\nu}$ is probably the most 
interesting channel in a long-term 
perspective.\footnote{~Concerning $B\to \mu^+\mu^-$, 
the curve reported in fig.~\ref{fig:conf} (right) has been 
obtained assuming a $10\%$ error on $f^2_{B_s}$, which we 
assume as a realistic estimate in a long-term 
perspective. This channel would become as clean as 
$K_L \to\pi^0 \nu\bar{\nu}$ (or equally sensitive to the scale
of $\cO_{\ell 1}$) if the error on $f^2_{B_s}$ were reduced 
below 5\%. }

\subsection{Non-leptonic decays}

In principle also non-leptonic $B$ and $K$ decays
could be used to put constrains on MFV operators: 
to obtain significant bounds it is necessary 
to disentangle the tiny contribution of electroweak 
operators (${\cal Q}_{7\ldots 10}$) from the dominant 
effects due to tree-level and gluon-penguin amplitudes 
(described by ${\cal Q}_{1\ldots 6}$). Penguin-dominated 
$B$ decays, such as $B \to K\pi$, seems the most promising 
candidates for this program. However, 
theoretical and experimental uncertainties 
are still quite large and do not allow to extract 
competitive bounds at the moment. 

Hadronic uncertainties are drastically reduced in 
CP-violating time-dependent asymmetries, such as 
$a^{CP}_{\Psi K}$. However, in the MFV scenario 
we do not expect significant effects on these 
observables due to the absence of new CP-violating phases. 
For instance, a prediction of the  MFV scenario
is that CP asymmetries in $B\to \phi K_S$ and  
$B\to \Psi K_S$ decays are approximately the same,
as in the SM.

\section{Yukawa interaction with two Higgs doublets}
\label{sect:2HD}

As anticipated, the two-Higgs-doublet (2HD) case is particularly
interesting since we can se\-pa\-rate the breaking of
${\rm SU}(3)^3_q$, induced by the $\yuk$, and
the breaking of ${\rm U}(1)_{\rm PQ}$. The latter
is usually invoked to forbid tree-level FCNCs, which
would naturally arise if $H_U$ and $H_D$ can couple to both
up- and down-type quarks. Indeed if $H_D$ has
${\rm U}(1)_{\rm PQ}$ charge opposite to $\bar D_R$, and
$H_U$ is neutral, the ${\rm U}(1)_{\rm PQ}$-invariant effective Yukawa
interaction is
\beq
\cL_{Y_0}  =   {\bar Q}_L \yuk_D D_R  H_D
+ {\bar Q}_L \yuk_U U_R  H_U
+ {\bar L}_L \yuk_E E_R  H_D {\rm ~+~h.c.}
\label{eq:LY2}
\eeq
In this framework the smallness of the $b$ quark and $\tau$ lepton masses is
naturally attributed to the smallness of $\langle
H_D\rangle/\langle H_U\rangle
= 1/\tan\beta$ and not to the corresponding Yukawa couplings. This implies,
in particular, that $\yuk_D \yuk^\dagger_D$ represents a
new non-negligible $(8,1,1)$ source of ${\rm SU}(3)^3_q$
breaking. Up to negligible terms suppressed by $m_{s,d}/m_b$,
this spurion can be written, in the basis ({\ref{eq:d-basis}), as
\beq
\yuk_D \yuk^\dagger_D \approx \frac{2 m^2_b \tan^2\beta}{v^2}  \Delta~, \qquad
\Delta = {\rm diag}(0,0,1)~.
\eeq

The ${\rm U}(1)_{\rm PQ}$ symmetry cannot be exact:
it has to be broken at least in the scalar potential
in order to avoid the presence of a massless pseudoscalar Higgs.
Coherently with our general definition
of minimal flavour-violating models, we shall assume that the
${\rm U}(1)_{\rm PQ}$ breaking does not induces new sources of
${\rm SU}(3)^3_q$ breaking. Despite this minimal assumption,
we can still induce an important modification on the Yukawa interaction,
allowing terms of the type
\beqa
 \epsilon_i {\bar Q}_L  (\yuk_D\yuk_D^\dagger)^{n_1} (\yuk_U\yuk_U^\dagger)^{n_2}
 (\yuk_D\yuk_D^\dagger)^{n_3} \yuk_D
 D_R  (H_U)^c~,
\label{eq:O_PCU} \\
 \epsilon_j {\bar Q}_L  (\yuk_D\yuk_D^\dagger)^{n_4} (\yuk_U\yuk_U^\dagger)^{n_5}
 (\yuk_D\yuk_D^\dagger)^{n_6} \yuk_U
 U_R  (H_D)^c~,
\label{eq:O_PCD}
\eeqa
where the $\epsilon_i$ denote generic ${\rm SU}(3)^3_q$-invariant
${\rm U}(1)_{\rm PQ}$-breaking
sources. Even if $\epsilon_i \ll 1 $, the product
$\epsilon_i \times  \tan\beta$ can be $\cO(1)$, inducing
$\cO(1)$ corrections to $\cL_{Y_0}$~\cite{HRS}.

Since the freedom to redefine the $\yuk$ has
already been used to eliminate higher-order operators
from $\cL_{Y_0}$, we cannot trivially ignore
 ${\rm U}(1)_{\rm PQ}$-breaking
operators with several powers of $\yuk$. However, we can
still perform the usual expansion in powers of
suppressed off-diagonal CKM elements.
To first non-trivial order, this leads to the following
modification of $D_R$ and $U_R$ Yukawa interactions
\beqa
\cL_{\epsilon {Y_D} }  &=&
  {\bar Q}_L \left( \epsilon_0 + \epsilon_1  \Delta + \epsilon_2
 \hatFC + \epsilon_3 \hatFC \Delta
 +\epsilon_4 \Delta \hatFC \right) \hatdyuk_d D_R (H_U)^c {\rm ~+~h.c.},
\label{eq:LYPQ_D} \\
\cL_{\epsilon {Y_U} }  &=&
  {\bar Q}_L \left( \epsilon'_0 + \epsilon'_1  \Delta + \epsilon'_2
 \hatFC + \epsilon'_3 \hatFC \Delta
 +\epsilon'_4 \Delta \hatFC \right) \hatV^\dagger \hatdyuk_u U_R (H_D)^c  {\rm ~+~h.c.},
\label{eq:LYPQ_U}
\eeqa
written in the basis ({\ref{eq:d-basis})\footnote{~The hat over
$V$ and $\lambda_i$ indicates that these quantities do not
satisfy the usual relations to quark masses and CKM angles.
We are working in a basis where the Yukawas in eq.~(\ref{eq:LY2}) are written
as $Y_D= \hatdyuk_d$, $Y_U= \hatV^\dagger \hatdyuk_u$, where
$\hatdyuk_{d,u}$ are diagonal matrices and $\hatV$ is a unitary matrix.
Also we define $(\hatFC )_{ij}= (\hatV^\dagger)_{i3}(\hatV)_{3j}$
when $i\ne j$, and equal to zero otherwise.
We assume that both $(\hatdyuk_u)_{33}$ and  $(\hatdyuk_d)_{33}$
are of $\cO(1)$ and reabsorb possible polynomials of these
variables in the definition of the $\epsilon_i$. The latter
are assumed to be real.} for the $\yuk$
and defining the Higgs doublets as
\be
H_D = \left( \ba{c} H_D^+ \\ H_D^0 \ea\right), \quad
H_U = \left( \ba{c} H_U^0 \\ H_U^- \ea\right), \quad
(H_D)^c = \left( \ba{c} -H_D^{0*} \\ H_D^- \ea\right), \quad
(H_U)^c = \left( \ba{c} -H_U^+ \\ H_U^{0*} \ea\right). \quad
\label{eq:Hc_conv}
\ee

As discussed in specific supersymmetric scenarios,
for $\epsilon_i \tan\beta = \cO(1)$ the ${\rm U}(1)_{\rm PQ}$-breaking
terms induce $\cO(1)$ corrections to down-type Yukawa
couplings~\cite{HRS}, CKM matrix elements~\cite{BRP}
and charged-Higgs couplings~\cite{CDGG,CGUW,DGG}, and
allows a sizeable FCNC coupling of down-type quarks to the
neutral Higgs fields~\cite{Babu,IR}.

\medskip

We shall proceed first with the diagonalization of down-type
mass terms generated by $\cL_{Y_0}+\cL_{\epsilon {Y_D} }$.
Since $\epsilon_i \tan\beta$ can be of order one,
we shall not make any expansion on these couplings.
On the other hand, we shall perform the usual
perturbative expansion in off-diagonal CKM elements.
Then the diagonalization of down-type  mass terms is obtained
by the rotation
\beq
D_L \to V_{D_L}D_L ~, \qquad D_R \to V_{D_R}D_R ~,
\label{eq:rot_2H}
\eeq
\beq
(V_{D_L})_{ij} =
\delta_{ij} - \frac{(\Sigma \diagyuk_d + \diagyuk_d \Sigma^\dagger)_{ij}}{
     \left(\diagyuk^2_d\right)_{ii} - \left(\diagyuk^2_d\right)_{jj} }
~,
\qquad
(V_{D_R})_{ij} =
\delta_{ij} - \frac{(\Sigma^\dagger \diagyuk_d + \diagyuk_d \Sigma)_{ij}}{
     \left(\diagyuk^2_d\right)_{ii} - \left(\diagyuk^2_d\right)_{jj} }
\eeq
where $\Sigma = \tan\beta(\epsilon_2 \hatFC  + \epsilon_3 \hatFC \Delta
 +\epsilon_4 \Delta \hatFC)\hatdyuk_d$  and $\diagyuk_d$
defines the mass eigenvalues via the relation
\beq
M_d = {\rm diag} \{ m_d,m_s,m_b\} = \langle H_D \rangle \diagyuk_d =  \langle H_D \rangle
\left[1+  (\epsilon_0 +\epsilon_1\Delta)\tan\beta \right] \hatdyuk_d~.
\eeq

The perturbations induced by $\cL_{\epsilon {Y_U} }$ on up-type mass
terms are very small, since they are suppressed by $\epsilon_i /\tan\beta$.
In the limit where we neglect such terms, the up-type mass matrix
remains unaffected by the ${\rm U}(1)_{\rm PQ}$-breaking
($\diagyuk_u = \hatdyuk_u$) and its diagonalization
is obtained by the usual rotation $U_L \to \hatV^\dagger U_L$.
However, because of the rotation of $D_L$ in eq.~(\ref{eq:rot_2H}),
$\hatV$ does not correspond anymore to the physical CKM matrix
($\realV$), which is defined by gauge charged-current interactions
to be $V=\hat V V_{D_L}$.
Keeping only the leading corrections, we find
\beq
\frac{ \hatV_{i3} }{\realV_{i3} } =
\frac{ \hatV_{3i} }{\realV_{3i} } = 1+r_V
~,  \qquad i\not=3;
\eeq
\beq
r_V \equiv \frac{(\epsilon_2+\epsilon_3)\tan\beta}{1 +
      (\epsilon_0+\epsilon_1-\epsilon_2-\epsilon_3)\tan\beta}~.
\eeq
In all other cases, to first order, $\realV_{ij}= \hatV_{ij}$.
Therefore, we also obtain
\beq
\lambda_t^2 \hatFC = (1+r_V) \left[ \lambda_{\rm FC}  + r_V (\lambda_{\rm FC}
- \Delta \lambda_{\rm FC} - \lambda_{\rm FC}\Delta) \right]~,
\eeq
where $\lambda_{\rm FC}$ is defined as in eq.~(\ref{eq:FC})
in terms of the physical CKM matrix and the top Yukawa coupling.

Because of the presence of two Higgs fields, with different vevs,
the diagonalization of the mass terms does not eliminate
scalar FCNC interactions. In the case of down-type
quarks, the effective FCNC coupling surviving after the
diagonalization can be written as 
\beq
\cL_{\rm FCNC}^{D}
=  \frac{1}{\cos\beta}\left[ 2 \sqrt{2} G_F \right]^{1/2}
{\bar D}_L \left( a_0 \lambda_{\rm FC} + a_1 \lambda_{\rm FC}  \Delta
     + a_2  \Delta \lambda_{\rm FC} \right) M_d D_R
    \left[ \frac{1}{\tan \beta} H_U^{0*}  -H_D^0  \right] {\rm +~h.c.},
\label{eq:LH_FCNC}
\eeq
where
\beqa
a_0 &=& \frac{  \epsilon_2 \tan\beta (1+r_V)^2 }{ \lambda_t^2
       \left[ 1+ \epsilon_0 \tan\beta\right]^2}~, \qquad\qquad
a_1 +a_0 = \frac{ r_V }{\lambda_t^2 \left[ 1+ (\epsilon_0+\epsilon_1)\tan\beta
\right]}~, \label{eq:ai} \no \\
a_2 -a_1 &=& \frac{(\epsilon_4-\epsilon_3 )\tan\beta}{ \lambda_t^2
\left[1+\epsilon_0 \tan\beta\right]
       \left[1 + (\epsilon_0+\epsilon_1-\epsilon_2-\epsilon_3)\tan\beta
\right]}~.
\eeqa
By construction, the neutral Higgs combination in eq.~(\ref{eq:LH_FCNC})
has zero vacuum expectation value
(and no Goldstone-boson component). In the limit where
$\tan\beta \gg 1$ and $M^2_{H^{\pm}} \gg M_W^2$, we can identify
it with the physical component of $H^0_D$, or with the combination
$[H^0 - i A^0]/\sqrt{2} \approx [\langle H_D \rangle-H_D^0]$
of heavy scalar ($H^0$) and pseudoscalar ($A^0$)
fields almost degenerate in mass
($M_{A^0} \approx M_{H^0} \approx M_{H^{\pm}}  $).

Also in the up sector an effective FCNC coupling of the type
${\bar U}^i_L {\bar U}^j_R [ H_U^{0}/{\tan \beta}-H_D^{0*}]$ arises:
its structure can easily be read from $\cL_{\epsilon {Y_U} }$
in eq.~(\ref{eq:LYPQ_U}), performing the rotation  $U_L \to \hatV^\dagger U_L$. However,
this term is less interesting than $\cL_{\rm FCNC}^{D}$
since in this case the $a'_i$  coefficients (defined
in analogy with the $a_i$) turn out to be $\cO(\epsilon'_i)$
and not of $\cO(\epsilon_i\tan\beta)$ as in eq.~(\ref{eq:ai}).

\medskip

Charged-current interactions with the physical Higgs boson
($H^+ \approx H^+_D$, in the large $\tan\beta$ limit)
also receive $\cO(1)$ corrections. Performing the diagonalization
and neglecting, among homogeneous terms, those suppressed by
$1/\tan\beta$, we can write
\beqa
\cL_{\rm H^+ } &=&  
    \frac{1}{\cos\beta}\left[ 2 \sqrt{2} G_F \right]^{1/2}
    \Big[ {\bar U}_L \left( b_0 \realV + b_1 \realV  \Delta
     + b_2  \Delta \realV  + b_3  \Delta \right) M_d D_R  \no \\
               & & \qquad + \frac{1}{\tan^2\beta}
    {\bar U}_R  M_u \left( b'_0 \realV + b'_1 \realV  \Delta
     + b'_2  \Delta \realV  + b'_3  \Delta \right) D_L \Big] H^+ {\rm ~+~h.c.}~,
\label{eq:LH_charged}
\eeqa
where the explicit expressions of $b_i$ and $b'_i$ in terms
of $\epsilon_i$ and $\epsilon'_i$ are
\beq
\ba{ll}
  b_0  = [1+ \epsilon_0 \tan\beta]^{-1}
& b_1 = -b_0 + [1 + (\epsilon_0+\epsilon_1-\epsilon_2-\epsilon_3)\tan\beta
]^{-1}  \\
  b_2 =  - \lambda_t^2 (a_0+a_2)
& b_3 = -(b_0+b_1+b_2)+ [1+ (\epsilon_0+\epsilon_1)\tan\beta]^{-1}  \\
  b'_0 = 1+ \epsilon'_0\tan\beta
& b'_1 = (\epsilon'_1 -\epsilon'_2 -\epsilon'_4)\tan\beta(1+r_V) \\
  b'_2 = \left[-r_V \epsilon'_1 +(1+r_V)(\epsilon'_2 +\epsilon'_3)\right]
\tan\beta \qquad
& b'_3 =-(b'_0+b'_1+b'_2) + 1 +(\epsilon'_0+\epsilon'_1)\tan\beta
\ea
\eeq
In this case the $\epsilon'$ coefficients cannot be neglected since they
induce $\cO(\epsilon' \tan\beta)$ modifications to the suppressed
${\bar U}_R D_L H^+$ coupling.

\section{FCNC transitions in the 2HD case}
\label{sect:2HD_FCNC}

In this scenario the determination of the effective
low-energy Hamiltonian relevant to FCNC processes
involves the following three steps:
\begin{itemize}
\item{} construction of the gauge-invariant basis of
dimension-six operators (suppressed by $\Lambda^{-2}$)
in terms of SM fields and two Higgs doublets;
\item{} breaking of ${\rm SU}(2)\times {\rm U}(1)_Y$ and
integration of the  $\cO(M_H^2)$ heavy Higgs fields;
\item{} integration of the $\cO(M_W^2)$ SM degrees
of freedom (top quark and electroweak gauge bosons).
\end{itemize}
These steps are well separated if we assume the
scale hierarchy $\Lambda \gg M_H \gg M_W$.
On the other hand, if $\Lambda \sim M_H$, the first
two steps can be joined, resembling the
one-Higgs-doublet scenario discussed before.
The only difference is that now, at large $\tan\beta$,
$\yuk_D$ is not negligible and this leads to enlarge
the basis of effective dimension-six operators.

In the rest of this section we shall first discuss the general
features of dimension-six operators, both for up-type and
down-type FCNC transitions. Then we shall analyse in detail
specific aspects of the second step, which characterize
the case $\Lambda \gg M_H$.

\subsection{Dimension-six FCNC operators with external down-type quarks}

Since $\yuk_D$ is not negligible, in this framework we can
in principle consider also $\Delta F=1$ operators built
in terms of the bilinear
\beq
{\bar D}_R \yuk_D^\dagger \yuk_U \yuk_U^\dagger \yuk_D D_R~,
\eeq
which has not been considered in section~\ref{secop}. However,
it is easy to realize that such terms are still very
suppressed.
Indeed in FCNC processes the two $\yuk_D$ cannot be both
contracted with an external $b$ quark:
this implies at least a $m_s/m_b =\cO(10^{-2})$ suppression with
respect to the first term in eq.~(\ref{eq:3FCNC}).
A similar argument holds also for the $\Delta F=2$
operator
\beq
({\bar D}_R \yuk_D^\dagger \yuk_U \yuk_U^\dagger Q_L)
({\bar Q}_L \yuk_U \yuk_U^\dagger \yuk_D  D_R)~.
\label{eq:BB_LR}
\eeq
It would not hold for $( {\bar D}_R \yuk_D^\dagger \yuk_U \yuk_U^\dagger Q_L)^2$,
but this structure is not ${\rm SU}(2)_L\times {\rm U}(1)_Y$ invariant
and therefore it can only arise at the level of dimension-8 operators
(the absence of this term is a clear example of the discriminating power
of our approach compared to a na\"\i ve analysis of dimension-six
four-quark operators).

The only new structure involving right-handed fields which cannot
be ignored is the $\Delta F=1$ scalar operator
\beq
({\bar D}_R \diagyuk_d \lambda_{\rm FC} Q_L) {\bar L}_L \diagyuk_\ell  E_R~.
\label{eq:Bll_LR}
\eeq
In the case of $B \to X_{s,d} \tau^+\tau^-$ transitions
its contribution is suppressed compared to vector and axial-vector ones
only by the product $(\diagyuk_d)_{33} (\diagyuk_\ell)_{33}$,
which may be of $\cO(1)$. Second, and more important,
in all $B_{s,d} \to \ell^+\ell^-$ amplitudes its effect
is suppressed only by $(\diagyuk^2_d)_{33}$ compared to the
axial-vector one which, in turn, is helicity suppressed
$[\cA(B_{s,d} \to \ell^+\ell^-)_{\rm axial} \propto 
m_\ell/m_b]$.\footnote{~As pointed out in ref.~\cite{Bob,Olive},
another helicity-suppressed observable sensitive to scalar 
operators, although rather challenging from the experimental point of view, 
is the lepton forward-backward asymmetry of 
$B \to  (\pi,K) \ell^+ \ell^-$ decays.}
We shall discuss the effect of this operator in more detail
later on. Here we simply note that, due to this term,
the bounds on $B_{s,d} \to \ell^+\ell^-$ decays reported in
table~\ref{tab:neg_lim} do not hold in a MFV scenario
with two Higgs doublets and large $\tan\beta$
(as explicitly shown in~\cite{Babu,IR,Bob,Bll}
in the framework of supersymmetry).

\medskip

In addition to the scalar operator in (\ref{eq:Bll_LR}),
in this framework we can generate a whole new set of
vector and axial-vector operators inserting the
spurion $\yuk_D \yuk_D^\dagger$ in the $\cO_i$ of section~\ref{secop}.
To be more precise, we can perform the replacement
\beq
{\bar Q}_L \lambda_{\rm FC}  Q_L \quad \longrightarrow \quad
{\bar Q}_L \Delta \lambda_{\rm FC}  Q_L \quad {\rm or} \quad
{\bar Q}_L \lambda_{\rm FC} \Delta  Q_L~.
\eeq
By this way we generate a new set of terms which affect
in the same way $b\to s$ and $b \to d$ transitions
and are negligible in $s\to d$ ones.
In the analysis of the unitarity triangle
these new terms do not spoil the fact that $|V_{ub}|$,
$\Delta M_{B_d}/\Delta M_{B_s}$ and $|a^{CP}_{J/\Psi K}|$
are insensitive to new-physics effects. However,
they allow us to modify independently $\Delta M_{B_d}$
and $\epsilon_K$. As a result, in this framework
there is no hope to distinguish the two solutions in
fig.~\ref{fig:CKMfit}. Similarly, the bounds in table~\ref{tab:neg_lim}
on $b\to (s,d)$ transitions obtained from $s \to d$ ones (and
viceversa) do not hold anymore.

Finally, concerning dipole operators, we note that in the absence
of ${\rm U}(1)_{\rm PQ}$ breaking the corresponding
dimension-six operators are obtained from ${\cal O}_{G1,F1}$
with the substitution $H \to H_D$. In this limit the
corresponding bounds on $\Lambda$ are completely
independent from $\tan\beta$. On the other hand, if we allow
for a large breaking of the ${\rm U}(1)_{\rm PQ}$ symmetry,
the bounds can receive receive $\cO(\epsilon\tan\beta)$
corrections. For $\epsilon\tan\beta > 1$,
when ${\rm U}(1)_{\rm PQ}$-breaking operators become dominant,
the bounds on their effective scales are
$(\epsilon\tan\beta)^{1/2}$ times larger than
those in table~\ref{tab:tab}.

\subsection{Dimension-six FCNC operators with external up-type quarks}

Although substantially enhanced with respect
to the one-Higgs-doublet case, FCNC transitions
with external up-type quarks remain well
below a realistic experimental sensitivity.
This is because in up-type transitions
the SM long-distance background is much
larger than in the down case. Then long-distance
contributions obscure possible tiny short-distance effects,
which in our framework are strongly suppressed by the CKM hierarchy.
To illustrate more clearly this point, we
discuss the case of
$D^0$--$\bar D^0$ mixing.

The basis ({\ref{eq:d-basis}) is not a convenient choice
to discuss FCNC operators with external up-type quarks.
In this case the most suitable basis is the one where
$\yuk_U$ is diagonal and the non-diagonal terms
are obtained from $\yuk_D \yuk_D^\dagger$. It is then
easy realize that the only non-negligible operator
contributing to $D^0$--$\bar D^0$ mixing is
\beq
 {\cal O}'_0 = \frac{1}{2}
  (\lambda_d)_{33}^4 \left( {\bar Q}_L  V \Delta V^\dagger
 \gamma_\mu  Q_L \right)^2~.
\label{eq:O0_D}
\eeq
This generates a contribution to $D^0$--$\bar D^0$ mass difference
which, adopting the usual $1/\Lambda^2$ normalization,
 can be written as
\beqa
\frac{\Delta M_D}{\Gamma_D} &=& \frac{\eta_D B_D F_D^2 M_D}{ 3  \Lambda^2 \Gamma_D }
    ( \lambda_d)_{33}^4
   \left| V_{td} V_{ts} \right|^2  \no\\
    & \approx &  10^{-3} \times
    \left(\frac{\sqrt{B_D} F_D}{200~\mbox{MeV}}\right)^2
    \left(\frac{1~\mbox{TeV}}{\Lambda}\right)^2 \left(\frac{m_b \tan\beta}{m_t}\right)^4~,
    \label{eq:DDmix}
\eeqa
where $\eta_D$, $B_D$ and $F_D$ are defined
in analogy to neutral $K$- and $B$-meson systems. The numerical coefficient
in (\ref{eq:DDmix}) is more than one order of magnitude
above the present experimental limit on $\Delta M_D/\Gamma_D$.
But, what is even more important, it is of the same order
of magnitude of long-distance contributions to
$\Delta M_D/\Gamma_D$~\cite{Petrov}. We then conclude that
this observables cannot be used to put significant
constraints on MFV models, even in the 2HD
case at large $\tan\beta$.

Similar arguments holds
for all rare charm decays which have a realistic
chance to be detected.
On the other hand, MFV expectations for cleaner observables,
such as $D\to K \nu\bar\nu$ or $D\to \mu^+ \mu^-$~\cite{Burdman},
are beyond the experimental reach, at least in the near future.

\subsection{Integration of the heavy Higgs fields}

The integration of the heavy Higgs fields leads to
calculable contributions to the Wilson coefficients
of the effective FCNC operators. These
can naturally be separated in two classes: those
which receive non-vanishing contributions already at
the tree-level and those which are generated only
at the one-loop level. 

The only structures generated at the tree level are scalar
operators of the type (\ref{eq:BB_LR}) or (\ref{eq:Bll_LR}), obtained
by integrating out the neutral Higgs fields in eq.~(\ref{eq:LH_FCNC}).
The leading $b_R$--$d_L$ components of these two terms are
\beqa
 \cH_{\rm tree}^{\Delta F=1} &=&
 \frac{a_0+a_1}{M_H^2}
 ({\bar b}_R \lambda_d \lambda_{\rm FC} d_L)({\bar \ell}_L \diagyuk_\ell \ell_R) {\rm ~+~h.c.} \\
 \cH_{\rm tree}^{\Delta F=2} &=&
 - \frac{(a_0+a_1)(a_0+a_2)}{  M_H^2}
 ({\bar b}_R \lambda_d \lambda_{\rm FC} d_L) ({\bar b}_L \lambda_{\rm FC} \lambda_d d_R)  {\rm ~+~h.c.},
\eeqa
where, as in section~\ref{sect:2HD}, we 
denote by $M_H$ the common mass of the
heavy Higgs fields in the large $\tan\beta$ limit. 
The most interesting effect induced by 
$\cH_{\rm tree}^{\Delta F=1}$ is a possible sizeable enhancement of 
$B\to \ell^+\ell^-$ rates, as discussed first in ref.~\cite{Babu}
in the context of supersymmetry. 
Taking into account the pseudoscalar contribution plus the ordinary
axial-vector one (induced by $Q_{10A}$) we can write
\beq
\BR(B_q \to \ell^+ \ell^-)
   = \frac{  f^2_{B_q } M_{B_q } \tau_{B_q } m_\ell^2 |C_{10A} V_{tb} V^*_{tq}|^2 
     \lambda_t^4}{32 \pi \Lambda_0^4}
    \left(1-\frac{4 m^2_\ell}{M^2_{B_q }}\right)^{1/2} \left[  (1+\delta_S)^2 +
       \left(1-\frac{4 m^2_\ell}{M^2_{B_q }}\right) \delta_S^2  \right]~,
\eeq
where
\beq
\delta_S = \frac{\pi \sw M^2_{B_q } \lambda_t^2(a_0+a_1) \tan^2\beta }{ \alpha M_H^2 C_{10A} }
\approx 27~(a_0+a_1) \left(\frac{\tan\beta}{50}\right)^2\left(\frac{500~\mbox{GeV}}{M_H}\right)^2~.
\eeq
The numerical result has been obtained by
setting $C_{10A}\approx 1$ (as in the SM) and using the $B_s$ mass.
Note that in this case there is no explicit dependence on the $b$-quark mass since
the $m_b$ factor of the Yukawa is cancelled by the $1/m_b$ in the
matrix element of the pseudoscalar current.

The $\Delta F=2$ term can induce a non-negligible contribution 
only to $\Delta M_{B_s}$. This can be written as
\beqa
\frac{\Delta M_{B_s}}{(\Delta M_{B_s})_{\rm SM}} &=& 1 -  B(\mu^2_b)m_b(\mu^2_b)m_s(\mu^2_b)
\frac{ 32 \pi \sw \lambda_t^4(a_0+a_1) (a_0+a_2)  \tan^2\beta}{\alpha M^2_H C_0(M_W^2)}
 \no \\
&\approx& 1 - 5~
(a_0+a_1)(a_0+a_2) \left(\frac{\tan\beta}{50}\right)^2 \left(\frac{500~\mbox{GeV}}{M_H}\right)^2~,
\label{eq:DMst2}
\eeqa
where $B(\mu^2_b) \approx 1$ takes into account the
ratio of scalar and vector matrix elements,
evaluated at a scale $\mu_b \approx m_b$. 
Eq.~(\ref{eq:DMst2}) generalizes the result of
ref.~\cite{Buras2} obtained in the framework of supersymmetry.
Note that, an important difference with ref.~\cite{Buras2}
arises already within supersymmetry, due to the fact that 
in eq.~(\ref{eq:DMst2}) we re-sum all the leading 
$\epsilon_i\tan\beta$ terms (hidden in the $a_i$ coefficients).
As pointed out in ref.~\cite{IR}, this makes an important numerical 
difference for $\epsilon_i\tan\beta =\cO(1)$.

\begin{figure}[t]
$$
\includegraphics[width=8cm]{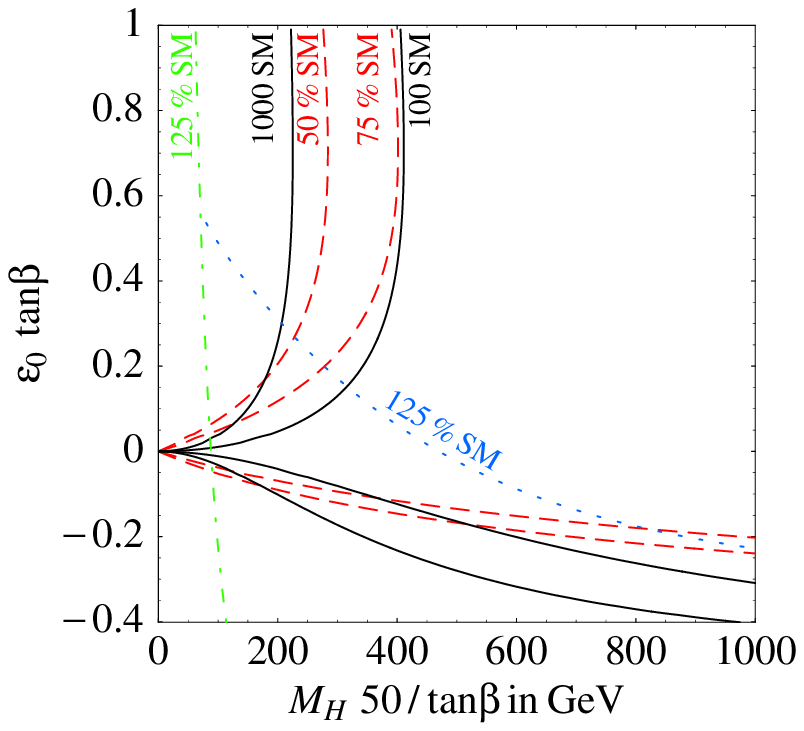}
\hspace{0.5cm}
\includegraphics[width=8cm]{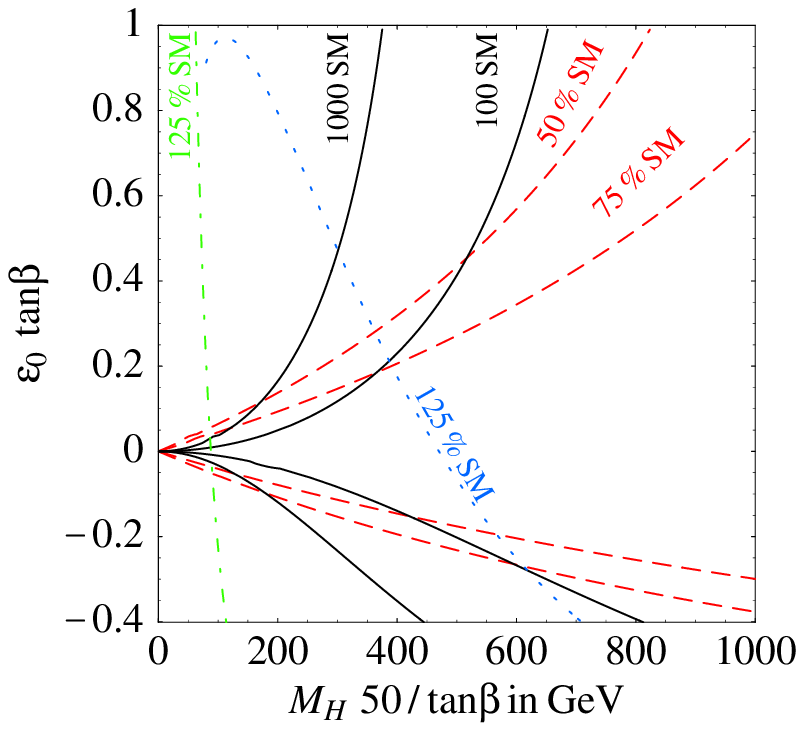}
$$
\caption[X]{\em {\bf Large $\tan\beta$ effects}:
$B\to \ell^+\bar\ell^-$ (continuous line), $\Delta M_{B_s}$ (dashed red line),
$B\to X_s \gamma$ (dotted blue line), $B\to  X \tau\bar{\nu}_\tau $ 
(dot-dashed green line). Left and right panels correspond to 
$\epsilon_2 = \epsilon_0$ and $\epsilon_2=-\epsilon_0$, respectively,
in addition to the constraints in eqs.~(\ref{eq:e_choice}).
\label{fig:tb}}
\end{figure}

The most interesting effects generated at the one-loop level are 
the contributions to dipole operators entering $B\to X_s \gamma$. 
Here we have non-vanishing results both from neutral- and charged-Higgs
exchange. Using the effective couplings derived in
section~\ref{sect:2HD} we find 
\beqa
\delta C_{7\gamma,8G}^{H^0}(M^2_W) 
  &=& - \frac{  (\epsilon_2 +\epsilon_3) \tan^3 \beta  }{ [1+(\epsilon_0+\epsilon_1) \tan\beta]^2 [1+(\epsilon_0+\epsilon_1-\epsilon_2-\epsilon_3)\tan\beta]}~\frac{ m_b^2(M^2_W)}{ 36 M_H^2}~, \\
 \delta C_{7\gamma,8G}^{H^\pm}(M^2_W) 
   &=&  \frac{1}{1+(\epsilon_0+\epsilon_1)\tan\beta} \left[  \frac{}{}
     1+(\epsilon_0'+\epsilon_2'+\epsilon_3')\tan\beta \right.  \no \\
   && \qquad\qquad \qquad\qquad \left. + \frac{(\epsilon_2+\epsilon_3)(\epsilon_2'+\epsilon_3'-\epsilon_1')\tan^2\beta}{1+(\epsilon_0+\epsilon_1-\epsilon_2-\epsilon_3)\tan\beta } \right] F_{7,8}^{(2)} \left( \frac{m^2_t}{M_H^2}\right)~,
  \label{eq:charged_bsg} \qquad 
\eeqa
where the functions $F_{7,8}^{(2)}(x)$ can be
found in appendix~\ref{app:bsg} [in our conventions 
$C_{7\gamma}^{\rm SM}(M^2_W)<0$].
These results extend those presented in ref.~\cite{DGG}. 
Even within supersymmetry we find a difference, since  
the neutral-Higgs-exchange amplitude and 
the $\epsilon_i^2\tan^2\beta$ terms in the charged-Higgs one 
(appearing only at the three-loop level in a pure diagrammatic approach)
have not been considered in ref.~\cite{DGG}.
However, since the neutral-Higgs contribution is numerically
small, the effect of these new corrections is limited. 
Explicit formul\ae\ for the supersymmetric case can be found 
in appendix~\ref{app:bsg}. 

\medskip

If all the $\epsilon_i$ are treated as free parameters,
the three constraints imposed by $B \to \ell^+\ell^-$,
$\Delta M_{B_s}$ and $B\to X_s \gamma$ are completely  
independent. On the other hand, in models where the  $\epsilon_i$ are
calculable these constraints can be combined, leading 
to interesting bounds on $M_H$ and $\tan\beta$. 
This happens for instance in supersymmetry (see appendix~\ref{app:bsg})
where, to a good approximation, 
\be
\epsilon_1  =   \epsilon_2~, \qquad 
\epsilon'_0 = - \epsilon_0~, \qquad 
\epsilon'_1 = - (\hatdyuk_b^2/\diagyuk_t^2) \epsilon_1~, \qquad  
\epsilon_3 = \epsilon_4 = \epsilon'_2=\epsilon'_3=\epsilon'_4  = 0~.
\label{eq:e_choice}
\ee
Motivated by this scenario, in fig.\fig{tb} we compare the three 
constraints employing eqs.~(\ref{eq:e_choice}),
supplemented by the additional (simplifying)
assumption $\epsilon_2= \epsilon_0$ or $\epsilon_2= - \epsilon_0$.
We use as variables $\epsilon_0 \tan\beta$ and $M_H/\tan\beta$
because all the effects, except the charged-Higgs amplitude in 
$B \to X_s \gamma$, depend only on these combinations. 
The $B\to X_s\gamma$ bound has been computed assuming $\tan\beta=50$;
it can be approximatively rescaled to other $\tan\beta$ values
taking into account that, in the limit where the (small) neutral-Higgs 
exchange amplitude is neglected, the horizontal scale (of this 
curve only) should be understood as $M_H$ instead of $(50/\tan\beta)M_H$.
For comparison, we have also shown the limit imposed by the tree-level
charged-Higgs exchange in $B\to X \tau  \bar{\nu}_\tau $ \cite{bXtau_ex},
which also depends only on $\epsilon_0 \tan\beta$ and $M_H/\tan\beta$
\cite{bXtau_th}: 
\be
\frac{\BR(B\to X \tau \bar{\nu}_\tau)}{\BR(B\to X \tau \bar{\nu}_\tau)_{\rm SM}} 
 = 1 - 1.2 (b_0 + b_1)
\bigg(\frac{m_\tau
\tan\beta}{m_H}\bigg)^2+
\bigg(\frac{m_\tau m_b (b_0+b_1)\tan^2\beta}{2 m_H^2} \bigg)^2~.
\ee
As can be noted, while the constraints from  $B\to X_s\gamma$
becomes weaker for large $\epsilon_0\tan\beta$ (especially for 
$\epsilon_2=\epsilon_0$), the opposite is true for 
$B_s\to \mu\bar\mu$ and $\Delta M_{B_s}$. As a result,
for any value of $\epsilon_0\tan\beta$
there is at least one FCNC process setting a bound on $M_H$ 
much stronger than the one imposed by $B\to X \tau  \bar{\nu}_\tau$. 

In the case of the precise measurements of  $B\to X \tau  \bar{\nu}_\tau$
and $B\to X_s\gamma$, we have shown in fig.\fig{tb} only one curve, 
corresponding to a maximal enhancement of $25\%$ with respect to the SM
value (close to the 90\%~C.L. limit in both cases).
In the case of $\Delta M_{B_s}$, where the new 
contribution leads to a decrease with respect to the SM value,
the two curves denotes $50\%$ and $75\%$ of the SM expectation. 
Due to theoretical uncertainties on the latter, the present 
90\%~C.L. limit is close to the $50\%$ curve. 
Finally, concerning  $\BR(B \to \ell^+ \ell^-)$, the two curves 
correspond to a possible enhancement of two or three orders of 
magnitude with respect to the SM value (for $\ell=e,\mu$). 
At the moment the best experimental bound is set by 
$\BR(B_s \to \mu^+ \mu^-)$ \cite{PDG}, whose 90\%~C.L. 
upper limit is about 900 times the SM value. Clearly, 
$B \to \ell^+ \ell^-$ decays still have a large discovery 
potential, for scenarios with large $\tan\beta$,
which will be explored in the future both at $e^+e^-$
$B$-factories and hadron colliders.

\section{Some concrete examples}
\label{sect:examples}

In this section we want to discuss the relation between the
MFV operators in our
effective theory approach and the flavour-violating
interactions in some commonly studied
extensions of the SM (partly motivated by the hierarchy problem).

\subsection{Supersymmetry}

Let us first consider the minimal supersymmetric extension of the SM,
with conserved $R$-parity and supersymmetry soft-breaking terms. To
understand its structure of flavour-violating couplings, it is convenient
to follow the same approach used in previous sections 
for the SM and the 2HD model. 
We then take the supersymmetric model as the low-energy effective theory, 
and we construct all flavour-invariant interactions that contain the 
spurion fields $Y_{U,D}$. An important difference with the non-supersymmetric 
case is that we find, beside the
ordinary Yukawa couplings, other renormalizable terms with non-trivial
flavour structure. Indeed, following the MFV rules, we can write 
the supersymmetry-breaking squark mass terms and the trilinear couplings as
\bea
{\tilde m}_{Q_L}^2 &=& {\tilde m}^2 \left( a_1 \identity 
+b_1 Y_U Y_U^\dagger +b_2 Y_D Y_D^\dagger 
+b_3 Y_D Y_D^\dagger Y_U Y_U^\dagger
+b_4 Y_U Y_U^\dagger Y_D Y_D^\dagger 
 \right)~,
\label{prima}\\
{\tilde m}_{U_R}^2 &=& {\tilde m}^2 \left( a_2 \identity 
+b_5 Y_U^\dagger Y_U  \right)~,\\
{\tilde m}_{D_R}^2 &=& {\tilde m}^2 \left( a_3 \identity 
+b_6 Y_D^\dagger Y_D  \right)~,\\
A_U &=& A\left( a_4 \identity 
+b_7 Y_D Y_D^\dagger  \right) Y_U~,\\
 A_D &=& A\left( a_5 \identity 
+b_8 Y_U Y_U^\dagger  \right) Y_D~.
\label{ultima}
\eea
Here $\tilde m$ and $A$ set the mass scale of the soft terms, $a_i$ and $b_i$
are unknown numerical coefficients, and $\identity$ is the $3\times 3$
identity matrix. 
In obtaining eqs.~(\ref{prima})--(\ref{ultima}), we have kept all 
independent flavour structures proportional to third-generation
Yukawa couplings (similarly to the $\epsilon_i$ of sect.~\ref{sect:2HD}, 
possible polynomials of $(\diagyuk_{u,d})_{33}$ can be 
reabsorbed in the definition of $a_i$ and $b_i$) 
but we have neglected contributions quadratic 
in the Yukawas of the first two families. 
When $\tan\beta$ is not too large and the bottom Yukawa coupling is small,
the terms quadratic in $Y_D$ can be dropped from 
eqs.~(\ref{prima})--(\ref{ultima}). 

The assumption of universality of soft masses and proportionality of 
trilinear terms corresponds to setting all $b_i$ coefficients to zero.
This hypothesis is not renormalization-group invariant, and it is usually
taken to hold at a high-energy scale $\Lambda$, whose value can be inferred
by knowledge of the specific mechanism of supersymmetry breaking.
Then the coefficients $b_i$ are generated by renormalization-group evolution,
and their typical size is $(1/4\pi)^2 \ln \Lambda^2/ {\tilde m}^2$.
For sufficiently large values of $\Lambda$, the effect can be significant.
The $b_i$ coefficients can also be generated by integrating out heavy states
at the cut-off scale $\Lambda$ (assuming that the high-energy dynamics
satisfies MFV). This mechanism can also lead to sizable effects, since 
the new contributions
to $b_i$ are not suppressed by powers of the scale $\Lambda$. This high-energy
sensitivity of the flavour structure in the soft terms~\cite{hkr} makes the
MFV assumption in supersymmetry rather tottery. In particular, in supergravity
where $\Lambda$ has to be identified with a scale close to the Planck mass,
MFV in the soft terms requires that all dynamics up to the gravitational
scale must satisfy MFV. This is quite possible but, from the low-energy point
of view,  it appears as a 
strong assumption. The situation improves in models with gauge-mediated
supersymmetry breaking~\cite{gm}, since the scale $\Lambda$ is identified
with the mass of hypothetical messenger particles. This mass scale is 
arbitrary, and it can be low enough
to reduce the region of high-energy sensitivity. In anomaly 
mediation~\cite{am} the problem is bypassed, since the soft terms are
driven towards universality by renormalization-group evolution, irrespectively
of their high-energy values.  

Using the soft terms in eqs.~(\ref{prima})--(\ref{ultima}), the physical
squark $6\times 6$ mass matrices, after electroweak breaking, are given by
\beq
{\tilde M}_U^2 = \pmatrix{
{\tilde m}_{Q_L}^2+Y_UY_U^\dagger v_U^2+\left( \frac{1}{2} -\frac{2}{3}\sw
\right) M_Z^2\cos 2\beta & \left( A_U -\mu Y_U \cot \beta \right) v_U \cr
 \left( A_U -\mu Y_U \cot \beta \right)^\dagger v_U & \!\!\!\!\!\!
{\tilde m}_{U_R}^2+Y_U^\dagger Y_U v_U^2+\frac{2}{3}\sw
 M_Z^2\cos 2\beta }~,
\label{mmup}
\eeq
\beq
{\tilde M}_D^2 = \pmatrix{
{\tilde m}_{Q_L}^2+Y_DY_D^\dagger v_D^2-\left( \frac{1}{2} -\frac{1}{3}\sw
\right) M_Z^2\cos 2\beta & \left( A_D -\mu Y_D \tan \beta \right) v_D \cr
 \left( A_D -\mu Y_D \tan \beta \right)^\dagger v_D & \!\!\!\!\!\!
{\tilde m}_{D_R}^2+Y_D^\dagger Y_D v_D^2-\frac{1}{3}\sw
 M_Z^2\cos 2\beta }~.
\label{mmdo}
\eeq
Here $\mu$ is the higgsino mass parameter and $v_{U,D}$
the two Higgs vacuum expectation values ($\tan\beta =v_U/v_D$).

The squark mass matrices are diagonalized by the unitary transformations
$T_{U,D}{\tilde M}_{U,D}^2 T_{U,D}^\dagger$. When all $b_i$ coefficients
in eqs.~(\ref{prima})--(\ref{ultima}) are negligible, in the basis
of eq.~(\ref{eq:d-basis}), the rotation
matrices are given by 
\beq
T_U=\pmatrix{ C_U V & S_U \cr -S_U V & C_U}~, ~~~~
T_D=\pmatrix{ C_D  & S_D \cr -S_D  & C_D}~,
\eeq
\beq
C_{U,D}={\rm diag} \left( \cos \theta_{U_i,D_i}\right)~, ~~~~
S_{U,D}={\rm diag} \left( \sin \theta_{U_i,D_i}\right)~,
\eeq
\beq
\tan 2 \theta_{U_i}= \frac{2(Aa_4-\mu \cot\beta)m_{u_i}}{(a_1-a_2)
{\tilde m}^2+\left( \frac{1}{2} -\frac{4}{3}\sw
\right) M_Z^2\cos 2\beta }~,
\eeq
\beq
\tan 2 \theta_{D_i}= \frac{2(Aa_5-\mu \tan\beta)m_{d_i}}{(a_1-a_3)
{\tilde m}^2-\left( \frac{1}{2} -\frac{2}{3}\sw
\right) M_Z^2\cos 2\beta }~.
\eeq
Here $m_{u_i}$ and $m_{d_i}$ are the corresponding tree-level quark masses.
In this case, the rotations in flavour space of squarks and quarks are 
the same, therefore tree-level flavour-changing transitions occur only 
in the charged current coupled to charginos, with angles equal to those of 
$V$. Notice that the left-right squark mixing does not modify the flavour
structure, although it complicates the expressions of the diagonalization
matrices.

When non-vanishing  $b_i$ coefficients are included in the soft terms,
the diagonalization of 
the mass matrices in eqs.~(\ref{mmup}) and (\ref{mmdo}) 
is, in general, more involved. However, 
if $\tan\beta$ is not large and the bottom
Yukawa coupling can be neglected, we can easily express the squark
rotations as
\beq
T_U=\pmatrix{ C_U' V & S_U' \cr -S_U' V & C_U'}, ~~~~
T_D=\pmatrix{ V  & 0 \cr 0  & \identity}~,
\eeq
\beq
C_{U}'={\rm diag} \left( \cos \theta_{U_i}'\right)~, ~~~~
S_{U}'={\rm diag} \left( \sin \theta_{U_i}'\right)~,
\eeq
\beq
\tan 2 \theta_{U_i}'= \frac{2(Aa_4-\mu \cot\beta)m_{u_i}}{(a_1-a_2)
{\tilde m}^2+(b_1-b_5) {\tilde m}^2   m_{u_i}^2/v_U^2 +
\left( \frac{1}{2} -\frac{4}{3}\sw
\right) M_Z^2\cos 2\beta }~.
\eeq
In this case, the rotation in flavour space of down-type quarks and
left down squarks is different, and tree-level flavour-changing transitions 
occur in neutral currents coupled to gluinos and neutralinos, 
with angles equal to those of $V$.

The next step is to integrate out the supersymmetric particles at one loop.
In the limit where the soft mass parameters are sufficiently larger than 
the Higgs vacuum expectation values, this step should be performed 
before considering the ${\rm SU}(2)_L$ breaking. As a result, 
one obtains a series of higher-dimensional ${\rm SU}(2)_L$-invariant
operators, whose leading (dimension-six) terms are those discussed 
in the previous sections. Following this approach, the coefficients 
of the MFV operators are computable in terms of supersymmetric
soft-breaking parameters.
In principle, one could be worried by the fact that the 
assumption ${\tilde m} \gg v_{U,D}$ is not necessarily 
a good approximation; however, it should be noticed that  
the leading ${\rm SU}(2)_L$-breaking 
effects associated with the third family do not spoil the structure 
of the operator basis discussed in the previous sections, since they 
can be reabsorbed by appropriate coefficient functions. 
Notice also that the integration of the supersymmetric degrees of freedom 
leads to a modification of the renormalizable operators and, in particular, 
of the Yukawa interaction. As a result, in an effective field theory 
with supersymmetric degrees of freedom, the relations between $Y_{U,D}$ 
and the physical quark masses and CKM angles are potentially modified. 
Since no large logarithms are involved, this effect is usually small, 
with the exception of the large $\tan\beta$ case,
discussed in detail in sect.~\ref{sect:2HD} and \ref{sect:2HD_FCNC}.

We stress that the MFV hypothesis does not imply that the physical squark 
masses are all equal, but the form of the mass splittings is tightly constrained,
see eqs.~(\ref{mmup}) and (\ref{mmdo}). One could think of a departure
from MFV, by assuming that the squark masses of the third generation are
different from those of the other two, as motivated by the requirement
of a reduced fine tuning~\cite{dg}. In this case, some SU(3) factors
of the flavour group are broken into SU(2). Then new flavour 
spurions with non-vanishing background values along the third-generation
component, in a basis in which neither $Y_U$ nor $Y_D$ is necessarily
diagonal, should be added in the effective-theory analysis. The flavour
effects in this scenario have been studied in ref.~\cite{ckn2}.
 
\subsection{Technicolour and extra dimensions}

In technicolour theories, the generation of fermion masses leads to a
chronic problem with flavour violation. In the absence of a completely
successful solution, the MFV effective-theory approach seems well suited
to compare experimental results with model-building attempts. Indeed 
the approach we are following in this paper is closely related to the 
one pioneered by Chivukula and Georgi in the 
framework of technicolour~\cite{Georgi}.

\medskip 

Models with a low quantum-gravity scale, based on extra-dimensional
scenarios, are also a natural arena for the MFV effective theory \cite{Dvali}, 
since the underlying theory is unknown and its low-energy limit becomes
strongly-interacting at a scale which is expected to be not much larger
than the weak scale. In some cases the effects in extra-dimensional models
are nevertheless computable. Tree level exchange of Kaluza-Klein
excitations of SM fermions~\cite{Aguila},\footnote{~We thank the authors of 
ref.~\cite{Aguila} for pointing out an inconsistency 
in the earlier version of this discussion.}
of the SM vector bosons, of the Higgs, and
of the graviton do not give rise to
significant flavour-violating effects. 
For instance, it is easy to realize that
the exchange of the Higgs excitations can sizeably affect only four-fermion
operators involving the top quark, which are not relevant for our
discussion.

As noticed in~\cite{pomarol}, large tree-level flavour-violating
effects mediated
by Kaluza-Klein excitations of gauge bosons appear in models
in which fermions are confined on different locations within a thick
brane where gauge forces can propagate. Such models have been proposed to give
a geometrical justification of the observed hierarchy of quarks and
lepton masses~\cite{arkschm}. The dangerous flavour-violating interactions are
caused by the non-trivial profile of the excited gauge modes along the
new dimensions, which leads to flavour-dependent couplings with the
fermions situated at different positions in the extra coordinates. 
These models violate
the MFV hypothesis, and indeed the limit on the mass of the first
Kaluza-Klein gauge-boson mode is extremely stringent, of the order 
of 5000~TeV~\cite{pomarol}.


\medskip  

At first sight it  appears quite problematic to impose MFV to a
quantum-gravity theory with fundamental scale at the TeV. 
This is because we cannot play with
a very high-energy scale to suppress unwanted flavour-violating
operators. The most realistic attempt so
far proposed makes use of the extra dimensions to solve the problem. It
is based on locality (flavour-violating interactions are confined on one
brane) and geometry (we live on a second brane, distant from the first one, 
and therefore we observe very
suppressed flavour-violating effective interactions)~\cite{sav}.
In this scheme, to generate Yukawa couplings, it is
necessary to have bulk fields, charged under the flavour group, which
mediate the information of flavour symmetry breaking between the two
branes. These bulk fields can also generate new flavour violating effects,
which respect the hypothesis of MFV by construction. In the context
of models with a warped extra dimension~\cite{RS1}, the emergence of
bulk fields with flavour charge from the request of MFV has a nice
interpretation through the AdS/CFT correspondence. Indeed, if we impose
a global flavour symmetry on the 4-dimensional CFT, we find
in the AdS picture, 5-dimensional bulk gauge fields for the
flavour group and also Yukawa couplings promoted to scalar bulk
fields~\cite{rattazzi}. Using this language, one can also understand the
relation between the usual flavour suppressions due to operators with
a very high-energy effective scale and suppression due to
extra-dimensional locality, since the
distance along the 5th coordinate is mapped into the renormalization
energy scale  of the 4-dimensional CFT.

Vector bosons of the SU(3)$^5$ flavour symmetry 
in the bulk generate, at low energies,
various effective operators. Their tree-level exchange, in the case of
a 5-dimensional theory, is particularly interesting since the
coefficients of these operators are finite and computable. 
In the exact flavour limit 
(with the Yukawa couplings set to zero), the strongest constraint comes
from the flavour-conserving  operator $\frac{1}{2}(\bar
L \gamma_\mu \tau^a L)^2$, whose coefficient is limited by present data
on muon decay to have an effective scale larger than
$\Lambda\circa{>}5\TeV$ at $99\%$ CL~\cite{BS}. 
Once we consider insertions of the scalar bulk fields
$Y_{U,D,E}$ (whose background values determine the Yukawa couplings),
then specific MFV operators are generated:
the $\Delta F=2$ operator ${\cal O}_{0}$  and
some combinations of the $\Delta F=1$ four-quark operators.
Operators giving rise to $b\to s\gamma$ are generated at one-loop
level.

\section{Conclusions}

The starting point of our analysis has been a general definition of MFV, in
an effective-theory approach.
The basic assumption of MFV is that the only source of SU(3)$^5$ flavour
breaking is given by the background value of scalar fields $Y$ which have
the same transformation properties under the flavour group as the ordinary
Yukawa couplings. The interaction Lagrangian of the MFV effective theory 
valid below the energy scale $\Lambda$ is 
constructed, in terms of the SM fields and the spurions $Y$, by assuming
flavour and CP invariance.

We believe that this definition of MFV leads to a realistic description
of the minimal effects in 
flavour physics almost necessarily present in any extension of the Standard
Model with non-trivial dynamics at the scale $\Lambda$. It would be
unrealistic to impose the more restrictive assumption that 
all higher-dimensional effective operators are flavour invariant and contain
only SM fields (and not $Y$), since SU(3)$^5$ flavour is not a symmetry of the 
SM. Our definition of MFV is consistent with the existence of Yukawa
couplings in the low-energy theory. Also, the more restrictive assumption 
indeed does not hold in known extensions of the SM, like supersymmetry.
On the other hand, for an effective theory derived from supersymmetry, 
our definition coincides with
the usual requirements of supersymmetric MFV: $R$-parity conservation,
flavour-universal soft scalar masses, and trilinear terms proportional to
the corresponding Yukawa couplings.
In the case of supersymmetry, the effective-theory language employed 
here is just a useful bookkeeping device,
but its use is especially suited for theories in which the dynamics at
the scale $\Lambda$ is not fully known or perturbatively calculable,
as in the case of technicolour or theories with low-scale quantum gravity.

If the effective theory is built from the SM with a single Higgs 
doublet, we have found that all new (non-negligible) 
flavour-violating effects involve a single flavour structure $\lambda_{\rm 
FC}$. We have then presented a general classification of dimension-6 effective
operators containing $\lambda_{\rm FC}$ and we have collected 
in table~\ref{tab:tab} the present bounds
on their coefficients $1/\Lambda^2$ (assuming the presence of a single 
operator at a time). 

The most stringent bound, coming from $B\to X_s \gamma$, is on the 
coefficient of the
magnetic operator ${\cal O}_{F_1}$,
whose scale $\Lambda$ is constrained to be larger than 12.4~TeV (for
constructive interference with the SM amplitude) or 9.3~TeV (for
destructive interference) at 99\% CL. A narrow range of $\Lambda$ around
3~TeV is allowed by the data (see fig.~\ref{fig:bsgfit}), 
corresponding to a decay amplitude
equal in size to the SM value but opposite in sign. Leaving aside this
fine-tuned case (or the case in which the contribution from
the chromomagnetic operator ${\cal O}_{G_1}$ compensates the effect from
${\cal O}_{F_1}$), this result shows that $B\to X_s \gamma$ poses a
highly non-trivial bound on $\Lambda$, especially for theories in which
this cut-off scale has to be identified with the electroweak scale,
as in many theories motivated by the hierarchy problem.

The second best bound after $B\to X_s \gamma$
is obtained, at present, by the CP-violating parameter $\epsilon_K$, 
which constrains the scale of the $\Delta F=2$ effective 
operator ${\cal O}_{0}$ to be larger than 5.0 or 6.4~TeV (for 
constructive or destructive interference with the SM amplitude).
This bound is determined within a global fit of the CKM 
unitarity triangle which, to a large extent, is not affected by 
the existence of the new operators. In particular, time-dependent 
CP-violating asymmetries in the $B$ system are not by themselves,
or when compared to tree-level processes, 
good probes of MFV models. If MFV is realized, deviations from SM 
expectations can be detected only by measuring the full strenght 
of FCNC transitions.

An important feature of the existence of a single flavour-violating
structure $\lambda_{\rm FC}$ is the possibly of relating predictions
for various FCNC processes both in the $K$ and in the $B$ system. In
table~\ref{tab:neg_lim} we show limits on some FCNC decays (stronger than the present
experimental bounds) derived from the MFV assumption and from data on
other processes. If future experimental searches give evidence for
violation of any of the bounds in table~\ref{tab:neg_lim},
not only will we discover
new physics, but we will also learn that its dynamics at the scale 
$\Lambda$ has a flavour structure beyond MFV. This will have important 
consequences for model building.

Moreover, the MFV interplay between
various processes in $K$ and $B$ physics allows an interesting comparison
between different experimental searches, as illustrated in figs.~3 and~4.
For instance, future experimental investigation of the process 
$B\to X_s \ell^+ \ell^-$ is expected to
improve the sensitivity on the coefficients of the operators
${\cal O}_{\ell_1 \ldots \ell_3}$, and ${\cal O}_{H_1, H_2}$. These
new results (in case of a negative search) will strengthen the MFV limits
presented in table~\ref{tab:neg_lim}. 
Because of its theoretical cleanliness, the process
$K_L \to \pi^0 \nu \bar \nu$ has an even bigger potential for setting
bounds on $\Lambda$, in the long-term future.

A significant difference in the analysis emerges if we construct the effective
theory starting from the SM with an enlarged Higgs sector containing two
scalar doublets. Since the bottom-quark Yukawa coupling can now be 
non-negligible and even of order unity (when $\tan\beta$ is large and
of order $m_t/m_b$), then it is possible to construct flavour-violating 
operators with a flavour structure different than $\lambda_{\rm FC}$.
We have presented a general procedure to include all new flavour-violating
effects, when $\tan\beta$ is large. These effects can be parametrized
by the eight (PQ-violating)  $\epsilon$ coefficients
defined in eqs.~(\ref{eq:LYPQ_U}) and ~(\ref{eq:LYPQ_D}),
which modify the flavour-violating neutral and charged currents coupled
to the Higgs bosons as given in eqs.~(\ref{eq:LH_FCNC}) and (\ref{eq:LH_charged}). 
New significant contributions
appear in several $B$ processes, with the most interesting effects in
$B\to \ell^+ \ell^-$ and $B\to X_s \gamma$. These two processes play
a complementary r\^ole in constraining the charged Higgs mass, in the
large $\tan\beta$ limit, see fig.~\ref{fig:tb}. Indeed, $B\to \ell^+ \ell^-$ 
is more constraining
at large values of $|\epsilon | \tan\beta$ while, for small 
$|\epsilon | \tan\beta$, 
$B\to X_s \gamma$ becomes more effective.

The two-Higgs doublet analysis can also be used in supersymmetric models
with large $\tan\beta$. In this case, 
the $\epsilon$ parameters are determined by a one-loop diagram
involving supersymmetric particles. Our procedure gives the exact result
to all orders in $\epsilon \tan\beta$, in the ${\rm SU}(2)_L$-symmetric limit,
therefore it effectively
resums the leading terms at any loop level. With this technique, we have
found two-loop contributions to $B\to X_s \gamma$ mediated by neutral
Higgs bosons and three-loop contributions mediated by the charged Higgs
boson, which were not known but that can be non-negligible in the large 
$\tan\beta$ limit.  

%
%

\section*{Acknowledgements}
We thank P. Gambino, U. Nierste, and R. Rattazzi for interesting 
discussions and  F. del Aguila, A.J.~Buras, A. Nelson, M. Perez-Victoria,
and J. Santiago for useful comments.

\appendix
\section*{Appendix}

\section{Higgs-mediated $\tan\beta$-enhanced contributions to
$B\to X_s \gamma$ in supersymmetry}
\label{app:bsg}
In this appendix we collect the formul\ae\ necessary to 
compute the $\epsilon_i \tan\beta$ corrections
to the heavy-Higgs (charged and neutral)  $B\to X_s \gamma$
amplitude in  supersymmetry. As is well known, these 
corrections do not vanish in the limit of heavy sparticles: 
using the general results of section~\ref{sect:2HD}, we can re-sum 
them to all orders in the ${\rm SU}(2)_L$-symmetric limit
($v^2 \ll {\tilde m}^2$).

The dimension-four non-holomorphic terms generated
at one loop by gluino and Higgsino diagrams \cite{HRS},
computed in the ${\rm SU}(2)_L$-symmetric limit, can be written as 
\bea
 \cL_{\epsilon Y}^{\rm SUSY} &=& - \frac{ 2 \alpha_{\rm s}\mu }{3\pi\mg } \left[
   H_2 \left( \frac{\msqL^2}{\mg^2}, \frac{\msdR^2}{\mg^2} \right) 
  {\bar Q}_L \yuk_D D_R  (H_U)^c 
 - H_2 \left( \frac{\msqL^2}{\mg^2}, \frac{\msuR^2}{\mg^2} \right)
  {\bar Q}_L \yuk_U U_R  (H_D)^c \right]
 \no \\  
   && - \frac{A}{16 \pi^2 \mu }  
   H_2 \left( \frac{\msqL^2}{\mu^2}, \frac{\msuR^2}{\mu^2} \right)
   {\bar Q}_L  \yuk_U\yuk_U^\dagger \yuk_D D_R  (H_U)^c   \no \\ 
   && + \frac{A}{16 \pi^2 \mu }  
  H_2 \left( \frac{\msqL^2}{\mu^2}, \frac{\msdR^2}{\mu^2} \right)
{\bar Q}_L \yuk_U  \yuk_D\yuk_D^\dagger U_R  (H_D)^c~+{\rm h.c.},
\label{eq:L_nonholo}
\eea
where
\be
H_2(x,y) = \frac{x\ln\, x}{(1-x)(x-y)} + \frac{y\ln\, y}{(1-y) (y-x)}
\ee
and, as usual, $\mu$ denotes the supersymmetric Higgs mass term
and $A$ the three-linear soft-breaking term.\footnote{The opposite sign 
between $(H_U)^c$ and $(H_D)^c$ terms in eq.~(\ref{eq:L_nonholo})
is due to the convention (\ref{eq:Hc_conv}).}
Comparing $\cL_{\epsilon Y}^{\rm SUSY}$ with the general 
expression in eqs.~(\ref{eq:LYPQ_D})--(\ref{eq:LYPQ_U}), 
it follows 
\bea
&& \epsilon_0 = - \frac{ 2 \alpha_{\rm s}\mu }{3\pi\mg } 
 H_2 \left( \frac{\msqL^2}{\mg^2}, \frac{\msdR^2}{\mg^2} \right)~, \qquad 
\epsilon_0' =  \frac{ 2 \alpha_{\rm s}\mu }{3\pi\mg } 
 H_2 \left( \frac{\msqL^2}{\mg^2}, \frac{\msuR^2}{\mg^2} \right)~, \no \\
&& \epsilon_2 = - \frac{A\lambda_t^2 }{16 \pi^2 \mu }  
 H_2 \left( \frac{\msqL^2}{\mu^2}, \frac{\msuR^2}{\mu^2} \right)~, \qquad 
 \epsilon_1  =  \epsilon_2~,   \no \\
&& \epsilon_1' = \frac{A \hatdyuk_b^2}{16 \pi^2 \mu }  
 H_2 \left( \frac{\msqL^2}{\mu^2}, \frac{\msdR^2}{\mu^2} \right)~, \qquad 
\epsilon_3 = \epsilon_4 = \epsilon'_2=\epsilon'_3=\epsilon'_4  = 0~,
\eea
where
\be
\hatdyuk_b= \frac{1}{\cos\beta}\left[ 2 \sqrt{2} G_F \right]^{1/2}
\frac{ m_b }{[1+(\epsilon_0+ \epsilon_2)\tan\beta]}~.
\ee 
In the limit where right-handed up and down squarks 
are degenerate, only two of these terms are independent
and we recover the relations (\ref{eq:e_choice}).

\begin{figure}[t]
$$
\includegraphics[width=12cm]{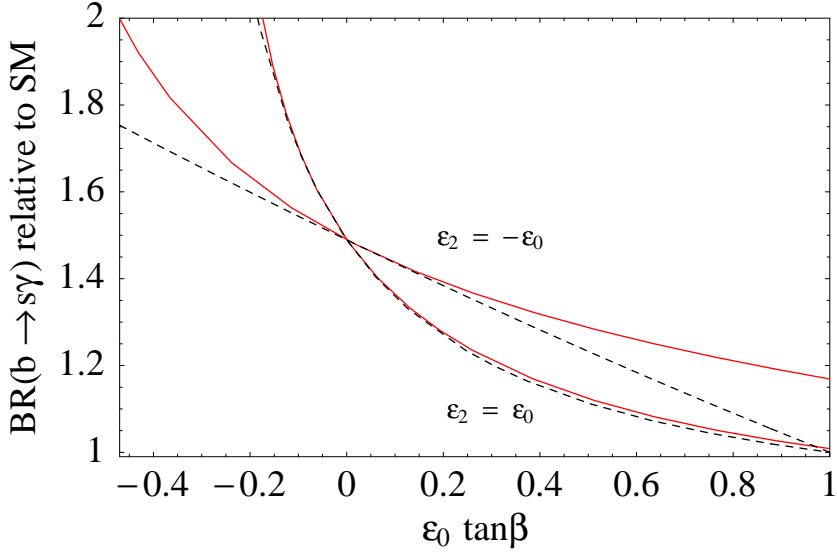}
$$
\caption[X]{\label{fig:bsg_susy}\em Deviations of 
$\BR(B\to X_s \gamma)$ with respect to the SM value
due to (supersymmetric) heavy-Higgs exchanges.
The two full (red) lines correspond to the full 
result in eq.~(\ref{eq:c7_susy_new}), for $\epsilon_2=\pm\epsilon_0$.
The dashed lines do not contain the new 
(higher-order) contributions identified in this paper (see text).  
All curves have been obtained for 
$M_H = 250$~GeV, $\tan\beta=50$ and $\epsilon'_0=-\epsilon_0$.}
\end{figure}

The supersymmetric $\epsilon_i$  induce four types 
of $\tan\beta$-enhanced corrections, which all 
affect the $B\to X_s \gamma$ amplitude: 
i) the modified relation between $m_b$ and 
the $b$-quark Yukawa coupling; ii) 
the modified relation between $m_t$ and 
the $\bar t_R d_L H^+$ vertex;  iii)
the FCNC ${\bar s}_L b_R H^0_D$ coupling; 
iv) the modification of $V_{ts}$. 
All these effects are taken into account to all orders 
when using the effective Lagrangians
(\ref{eq:LH_FCNC}) and (\ref{eq:LH_charged})
to compute the Higgs couplings. 
Computing both charged- and neutral-Higgs 
exchange amplitudes, at one loop, with these effective 
vertices, we find
\beqa
 C_{7\gamma,8G}^{H}(\muw)  &=& 
 \frac{1}{1+(\epsilon_0  +  \epsilon_2)\tan\beta}\left[ 
 1 + \epsilon'_0  \tan\beta
  - \frac{  \epsilon_2\epsilon_1' \tan^2\beta}{  1+ \epsilon_0\tan\beta } \right]  
F_{7,8}^{(2)} \left( \frac{m^2_t(\muw)}{M_H^2}\right) \no \\
  &-&  \frac{  \epsilon_2 \tan^3 \beta }{
  [1+(\epsilon_0+ \epsilon_2)\tan\beta]^2 [1+\epsilon_0\tan\beta]}~
\frac{ m_b^2(\muw)}{ 36 M_H^2}~, \qquad 
\label{eq:c7_susy_new}
\eeqa
where 
\bea
F_7^{(2)}(y)&=&\frac{y(3-5y)}{12(y-1)^2}+\frac{y(3y-2)}{6(y-1)^3}\ln y~, \\
F_8^{(2)}(y)&=&\frac{y(3-y)}{4(y-1)^2}-\frac{y}{2(y-1)^3}\ln y~.
\eea

The contributions which have not been considered before in the literature, 
namely the $\tan^2\beta$ term in the charged-Higgs part
and the full neutral-Higgs amplitude, are numerically 
rather suppressed, except for large $\epsilon \tan\beta$. 
As illustrated in figure~\ref{fig:bsg_susy},
these contributions are essentially negligible, 
independently of $\epsilon_0 \tan\beta$, if 
$\epsilon_2 \approx \epsilon_0$. On the other hand, 
sizable effects arise if $\epsilon_2 \approx  - \epsilon_0$. 
In particular, for $\epsilon'_0 \tan\beta \approx - \epsilon_0 \tan\beta \approx - 1$, 
where the leading term vanishes, these higher-order contributions 
could even become the dominant non-standard effect.

\section{The ${\cal Q}_i$ basis}
\label{app:qi}

Quark-lepton currents:
\beqa
{\cal Q}_{\nu\bar\nu}  &=& \bar d_i \gamma_\mu (1-\gamma_5) d_j ~  \bar \nu \gamma_\mu (1-\gamma_5) \nu \no \\
{\cal Q}_{10A}  &=& \bar d_i \gamma_\mu (1-\gamma_5) d_j ~  \bar \ell \gamma_\mu \gamma_5 \ell \no \\
{\cal Q}_{9V}   &=& \bar d_i \gamma_\mu (1-\gamma_5) d_j ~  \bar \ell \gamma_\mu \ell
\eeqa
Non-leptonic electroweak operators:
\beqa
{\cal Q}_7 &=& \bar d^\alpha_i \gamma_\mu (1-\gamma_5) d^\alpha_j
   ~\sum_q e_q  \bar q^\beta \gamma_\mu (1+\gamma_5) q^\beta \no \\
{\cal Q}_8 &=& \bar d^\alpha_i \gamma_\mu (1-\gamma_5) d^\beta_j
   ~\sum_q e_q  \bar q^\beta \gamma_\mu (1+\gamma_5) q^\alpha \no \\
{\cal Q}_9 &=& \bar d^\alpha_i \gamma_\mu (1-\gamma_5) d^\alpha_j
   ~\sum_q e_q  \bar q^\beta \gamma_\mu (1-\gamma_5) q^\beta \no \\
{\cal Q}_{10} &=& \bar d^\alpha_i \gamma_\mu (1-\gamma_5) d^\beta_j
   ~\sum_q e_q  \bar q^\beta \gamma_\mu (1-\gamma_5) q^\alpha
\eeqa
Dipole operators:
\beqa
{\cal Q}_{7\gamma} &=& \frac{1}{g^2}  m_{d_i} \bar d_i ( 1-\gamma_5)  \sigma_{\mu\nu}  d_j (e F_{\mu\nu})
 \no \\
{\cal Q}_{8G} &=& \frac{1}{g^2}  m_{d_i} \bar d_i ( 1-\gamma_5)  \sigma_{\mu\nu} T^a  d_j  (g_s G^a_{\mu\nu})
\eeqa

\frenchspacing
\footnotesize
\begin{multicols}{2}

\end{multicols}

\end{document}